\documentclass[journal]{IEEEtran}
\usepackage[utf8]{inputenc}

\usepackage{xcolor}
\usepackage{color, soul}
\usepackage{enumerate}
\usepackage[shortlabels]{enumitem}
\usepackage{url}

\usepackage{hyperref}
\usepackage{cite}
\usepackage{bm, bbm} 
\usepackage{mathtools}
\usepackage{amsmath, amssymb, amsthm}
\usepackage{nicefrac}
\usepackage{empheq}

\usepackage{booktabs}
\usepackage{graphicx}
\usepackage{multirow}
\usepackage{units}
\usepackage{siunitx}
\usepackage{enumitem}

\ifCLASSOPTIONcompsoc
\usepackage[caption=false, font=normalsize, labelfont=sf, textfont=sf]{subfig}
\else
\usepackage[caption=false, font=footnotesize]{subfig}
\fi

\usepackage{hyperref}
\usepackage{svg}

\usepackage{amsmath,amsfonts,amsthm,amssymb}
\usepackage[bb=boondox]{mathalfa}
\usepackage[algoruled]{algorithm2e}
\usepackage{array}

\usepackage[noabbrev,capitalize]{cleveref}

\crefname{equation}{}{}
\Crefname{equation}{}{}

\theoremstyle{definition} 

\theoremstyle{plain} 

\theoremstyle{remark} 
\newtheorem{remark}{Remark}

\usepackage{tikz}

\newenvironment{ldescription}[1]
  {\begin{list}{}%
   {\renewcommand\makelabel[1]{##1\hfill}%
   \setlength\itemsep{0pt}%
   \settowidth\labelwidth{\makelabel{#1}}%
   \setlength\leftmargin{\labelwidth}
   \addtolength\leftmargin{\labelsep}}}
  {\end{list}}

\newcommand{\set}[1]{\mathcal{#1}} 

\DeclareMathOperator{\cvar}{CVaR}

\DeclareMathOperator{\orr}{ORR}

\DeclareMathOperator{\ptdf}{PTDF}
\DeclareMathOperator{\lodf}{LODF}
\DeclareMathOperator{\eens}{EENS}

\makeatletter
\newcommand{\pushright}[1]{\ifmeasuring@#1\else\omit\hfill$\displaystyle#1$\fi\ignorespaces}
\newcommand{\pushleft}[1]{\ifmeasuring@#1\else\omit$\displaystyle#1$\hfill\fi\ignorespaces}
\makeatother

\newcolumntype{C}[1]{>{\centering\arraybackslash}p{#1}} 

\title{Risk-Aware Dimensioning and Procurement\\ of Contingency Reserve}
\author{Robert Mieth, Yury Dvorkin, and Miguel A. Ortega-Vazquez \vspace{-4mm}}

\begin{document}

\pagestyle{empty}
\bstctlcite{IEEE:BSTcontrol} 

\maketitle

\begin{abstract}

Current contingency reserve criteria ignore the likelihood of individual contingencies  and, thus,  their impact on  system reliability and risk.  
This paper develops an iterative approach, inspired by the current security-constrained unit commitment (SCUC) practice, enabling  system operators to determine risk-cognizant contingency reserve requirements and their allocation with minimal alterations to the current SCUC practice.
The proposed approach uses generator and transmission system reliability models, including failure-to-synchronize and adverse conditions, to compute contingency probabilities. {\color{black}These probabilities are then used to inform a reliability assessment of the system using a suitable risk metric, i.e., expected energy not served.} By iteratively learning the response of generators to post-contingency states within the SCUC we ensure reserve deliverability in this risk-assessment.  
The effectiveness of the proposed approach is demonstrated using the Grid Modernization Lab Consortium update of the Reliability Test System.

\end{abstract}

{\color{black}

\section*{Nomenclature}

\newcommand{\longestitem}{$C_{0,o,g}$}

\noindent\textit{Sets:}
\begin{ldescription}{\longestitem}
\item [$\mathcal{C}$] Set of contingencies
\item [$\mathcal{C^G}$] Set of contingencies with generator outages
\item [$\mathcal{C^L}$] Set of contingencies with line outages
\item [$\mathcal{C_\alpha}$] Set of worst-case contingencies at risk-level $\alpha$
\item [$\mathcal{G}$] Set of conventional generators
\item [$\mathcal{G}^c$] Set of generators with outage during contingency $c$
\item [$\mathcal{G}^{\text{FS}/\overline{\text{FS}}}$] Set of generators with/without fast-start ability
\item [$\mathcal L$ ] Set of lines 
\item [$\mathcal{L}^c$ ] Set of lines with outage during contingency $c$
\item [$\mathcal N$ ] Set of nodes 
\item [$\mathcal O$ ] Set of segments for operating cost curves of conventional generators 
\item [$\mathcal T$ ] Set of time periods
\item [$\mathcal W$ ] Set of RES (wind) generators
\item [$\mathcal{W}^D$ ] Set of dispatchable RES (wind) generators
\end{ldescription}

\vspace{0.5em}
\noindent\textit{Variables:}
\begin{ldescription}{\longestitem}
\item [$f_{ij,t}$] (Active) Power flow in line $ij$ at time $t$
\item [$p_{g,t}$] Power output for  generator $g$ at time $t$
\item [$r_{g,t}$] Total reserve of generator $g$ at time $t$
\item [$\hat{r}_{g,t,c}$] Called reserve activation from generator $g$ at time $t$ during contingency $c$
\item [$r^{NS}_{g,t}$] Non-spinning reserve of generator $g$ at time $t$
\item [$r^S_{g,t}$] Spinning reserve of generator $g$ at time $t$
\item [$r^{tot}_{t}$] Total reserve of all $g \in\mathcal G$ at time $t$
\item [$s_{i,t,c}^{\text{LS}}$] Load shedding at bus $i$ at time $t$ during contingency~$c$
\item [$s_{ij,t,c}^{+/-}$] Positive/negative overload on line $ij$ at time $t$ during contingency~$c$
\item [$t_{g,t}$] Auxiliary variable to compute the operating cost of generator $g$ at time $t$
\item [$u_{g,t}$] On/off commitment variable for  generator $g$ at time $t$
\item [$v_{g,t}$] Start-up binary variable for  generator $g$ at time $t$
\item [$w_{g,t}$] Shutdown binary variable for  generator $g$ at time $t$
\item [$\beta{g,t,c}$] Reserve activation factor of generator $g$ at time $t$ during contingency $c$
\item [$\theta_{i,t}$] Voltage angle at node $i$ at time $t$
\item [$\pi_c$] Probability of contingency $c$
\item [$\hat{\pi}_c^{\alpha}$] Adjusted conditional probability of contingency $c$ for risk-level $\alpha$
\item [$\Delta f_{ij,t,c}$] Change of line flow in line $ij$ at $t$ during during contingency $c$
\end{ldescription}

\vspace{0.5em}
\noindent\textit{Parameters:}
\begin{ldescription}{\longestitem}
\item [$\overline{p}_{w,t}$] Forecast wind production of wind farm $w$ at time $t$
\item [$q_g$] Forced outage rate (FOR) of generator $g$
\item [$q_g^{N/A}$] FOR of generator $g$ during normal/adverse conditions
\item [$q_g^s$] Failure-to-synchronize rate of generator $g$
\item [$B_{ij}$] Susceptance of line $ij$  
\item [$C_{g}^{0}$] No-load cost coefficient for  generator $g$  
\item [$C_{1,o,g}$] Linear cost coefficient of the operating cost for generator $g$ in cost segment $o$  
\item [$C_{0,o,g}$] Constant cost coefficient of the operating cost for generator $g$ in cost segment $o$  
\item [$C^{\text{LS}}$] Value of lost load
\item [$C^{\text{overl}}$] Penalty for line overload
\item [$C^{SU}_g$ ] Start-up cost of generator $g$ 
\item [$C^{SD}_g$ ] Shutdown cost of generator $g$ 
\item [$D_{i,t}$] Demand at node $i$ and time $t$ 
\item [$DT_g$] Minimum downtime (off) of generator $g$ 
\item [$P_{g}^{\max}$] Maximum power output for generator $g$  
\item [$P_{g}^{\min}$] Minimum power output for generator $g$  
\item [$f^{E}_{ij}$] Emergency maximum power flow limit in line $ij$
\item [$f^{\max}_{ij}$] Normal maximum power flow limit in line $ij$
\item [$F_g^A$] Share of outages of $g$ during adverse conditions
\item [$H_g$] Number of historical data samples for generator $g$
\item [$H_g^A$] Number of historical data samples for generator $g$ during adverse conditions
\item [$R^{D}$] Fraction of demand to be provided as reserve 
\item [$R^{S}$] Fraction of the total reserve that must be provided as spinning reserve
\item [$R^{10}_{g}$] 10-min ramp rate for generator $g$  
\item [$R^{60}_{g}$] 60-min ramp rate for generator $g$  
\item [$U_{g/ij,t}$] Probability that generator $g$/line $ij$ is unavailable
\item [$UT_g$] Minimum uptime (on) of generator $g$ 
\item [$\lambda$] Memory decay parameter for learning $\beta_{g,t,c}$
\end{ldescription}

\vspace{0.5em}
\noindent\textit{Additional symbols:}
\begin{ldescription}{\longestitem}
\item [$\mathbb{1}_{x\in\set{X}}$] Indicator function; 1 if $x\in\set{X}$, 0 else
\item [$\cdot^*$] Optimal value obtained from previous SCUC run
\end{ldescription}
}

\section{Introduction}

Reliable power system operation requires procurement of contingency reserves to respond to any unplanned outages of generation or transmission equipment.
Current approaches determine these requirements using deterministic security margins that (i) ignore the likelihood of potential contingencies and (ii) are defined in terms of system-wide or zonal quantities that trivialize  deliverability of the scheduled reserves in post-contingency system states. 
As a result, these methods are unable to trade-off the \textit{risk} of potential nodal power balance or flow limit violations against the cost of reserve provision and allocation.
Integrating contingency analyses and risk-based calculations for power system short-term planning purposes -- typically centered around solving an instance of the  security-constrained unit commitment problem (SCUC) -- is computationally demanding.
This paper develops an iterative approach, inspired by the current SCUC practice, that enables system operators to determine dynamic contingency reserve requirements and their allocation in a risk-aware manner with minimal alterations to the current SCUC practice.

Current practices for contingency reserve provision are adapted from reliability security standards, e.g., in the U.S. from the North American Electric Reliability Corporation (NERC) \cite{nercstandards}. 
Here, the minimum amount of reserve is set to comply with  given standards (e.g., BAL-002-2 for the U.S.) \cite{bal002}, that is  to withstand the most severe single contingency (i.e., ``N-1 criterion'').
U.S. system operators adhere to this minimal requirement with some modifications or extensions, which are typically static policies adjusted to a desired level of security in the system \cite{epri_reserve_dimensioning}.
For example, the California Independent System Operator (CAISO) and the Electric Reliability Council of Texas (ERCOT) require the total contingency reserve to cover the largest credible contingency and additionally constrain how different types of reserves (e.g., spinning and non-spinning, demand-side resources) contribute to the total amount \cite{caiso_ancillary_services,ercotNPRR863}.
Similarly, 
Independent System Operator New England (ISO-NE) and the New York Independent System Operator (NYISO) procure spinning and non-spinning contingency reserves to cover the largest contingency  within 10 minutes. Additional 30-minute reserves must cover another \unit[50]{\%} of the second largest contingency at ISO-NE, or together with the 10-minute reserves account for \unit[150]{\%} of the largest contingency at NYISO. See \cite{isone_operating_reserve_and_regulation,nyiso2020manual2}.

Although such minimal reserve requirements should cover the worst-case contingency (and, thus, implicitly less severe contingencies), they are typically scheduled by optimizing a pre-contingency system state, i.e., assuming normal operation alone. Therefore, they may not be deliverable in post-contingency system states due to system limits (e.g.,  congestion). 
Some ISOs ensure reserve deliverability implicitly by enforcing zonal reserve criteria and approximate inter-zonal exchange capacities \cite{wang2014dynamic}.
However, statically defined zones with typically long update intervals, e.g., yearly or quarterly \cite{wang2014dynamic}, may not reflect the actual system state and scheduled reserves may not be deliverable due to interzonal congestion. Additionally, pre-defined zonal reserve requirements produce suboptimal generator dispatch solutions \cite{arroyo2005energy}.
Ideally, post-contingency reserve deliverability should be  endogenous to  the SCUC optimization. 
However, this would lead to  computationally intractability, even if only ``N-1'' outages are considered \cite{capitanescu2011state}.
To alleviate this complexity in practice, heuristics and approximate approaches are often used.
For example, \cite{chen2013incorporating} describes a simplified security-constrained economic dispatch (SCED) formulation with fixed zonal load shift factors to model power flow changes caused by severe outages.
However, this approach relies on fixed zones and ignores potential intrazonal congestion. 
{\color{black} 
A similar approach is implemented by CAISO \cite{caiso_generator_cont_action_scheme} and uses  predefined generation distribution factors, computed from the technical parameters of generators providing frequency response services, to estimate post-contingency power flows.}
While  \cite{chen2013incorporating,caiso_generator_cont_action_scheme} approximate  post-contingency power flows, the underlying shift factors do not consider individual post-contingency system states explicitly, which reduce their applicability for varying system conditions. 
On the other hand, reserve activation factors, see e.g., \cite{singhal2017reserve,singhal2018data}, that model how scheduled reserves are called upon under various post-contingency states  improve reserve deliverability and reduce the cost of emergency corrective actions. 

While considering transmission constraints and post-contingency system states,  \cite{madani2016constraint,street2010contingency,chen2013incorporating,caiso_generator_cont_action_scheme,singhal2017reserve,singhal2018data} neglects the probability of generator and transmission contingencies and therefore cannot assess risk imposed by the contingencies using cost-benefit analyses.
To internalize the trade-off between system reliability and the cost of reserve provision, \cite{chattopadhyay2002unit,bouffard2004electricity} introduce approximations of reliability metrics, i.e., loss-of-load probability (LOLP) and expected energy not served (EENS), into the unit commitment problem.  In \cite{chattopadhyay2002unit,bouffard2004electricity}, instead of meeting a fixed total requirement, contingency reserves are scheduled with respect to a target LOLP or EENS.  
Alternatively, avoiding the selection of fixed LOLP or EENS targets, the cost of reserve provision and the expected cost of contingencies can be co-optimized by considering the value of lost load (VOLL) \cite{kariuki1996evaluation} in the objective of the SCUC formulation.
For example, \cite{ortega2006optimising} proposes a piecewise linear approximation of the EENS to obtain a computationally tractable solution, and \cite{bouffard2005market} applies scenario-based stochastic programming to co-optimize pre-contingency and expected post-contingency costs.
However,  \cite{chattopadhyay2002unit,bouffard2004electricity,kariuki1996evaluation,ortega2006optimising,bouffard2005market} require significant modifications to the current SCUC practice in the industry, which is an adoption barrier. On the other hand,  \cite{ortega2007optimizing} proposes an offline optimization of reserve requirements, which could then be enforced in the SCUC optimization with minimal alterations, by jointly minimizing the system operating cost of a reduced system model and the penalized EENS. This method has also been extended to account for the failure of generators  to synchronize \cite{ortega2008optimising} and wind uncertainty \cite{ortega2008estimating}. 

The notable limitation is that  \cite{chattopadhyay2002unit,bouffard2004electricity,kariuki1996evaluation,ortega2007optimizing,ortega2006optimising,bouffard2005market,ortega2008optimising,ortega2008estimating} do not consider  deliverability of scheduled reserves, i.e., they do not consider transmission systems constraints in either pre- and post-contingency states. On the other hand,  \cite{fernandez2016probabilistic,guerrero2018incorporating} 
model a network-constrained SCUC, which  minimizes the system cost and expected cost of load shedding, and endogenously compute  probabilities of contingencies as a function of commitment decisions.  
While this approach addresses reserve deliverability and internalizes contingency risks, it significantly alters SCUC computation and introduces numerous auxiliary binary and non-binary variables that may obstruct computation for large-scale networks.

This paper proposes to account for risk-aware reserve dimensioning, allocation and deliverability in the close-to-reality SCUC framework by learning risk-aware reserve activation factors. These factors can be learned iteratively and capture the post-contingency system states and their reliability. 
Once obtained such factors can be used to effectively approximate post-contingency power flows in pre-contingency system optimization with minimal alterations to the original SCUC practice. 
{\color{black}%
Relative to the previous work that also discusses the application of reserve activation factors, e.g., \cite{singhal2017reserve,singhal2018data}, the proposed approach considers not only generator outages but also transmission  outages in its risk-informed decision making, and formalize worst-case reserve deliverability constraints for a given measure of risk. Furthermore,instead of enforcing all post-contingency states in the SCUC formulation, the proposed approach uses conditional value-at-risk, as a risk measure, to select consistently and equitably 
worst-case contingencies based on their system impacts.}

\section{Model Formulation}

\begin{figure}
    \centering
    \includegraphics[width=0.95\linewidth]{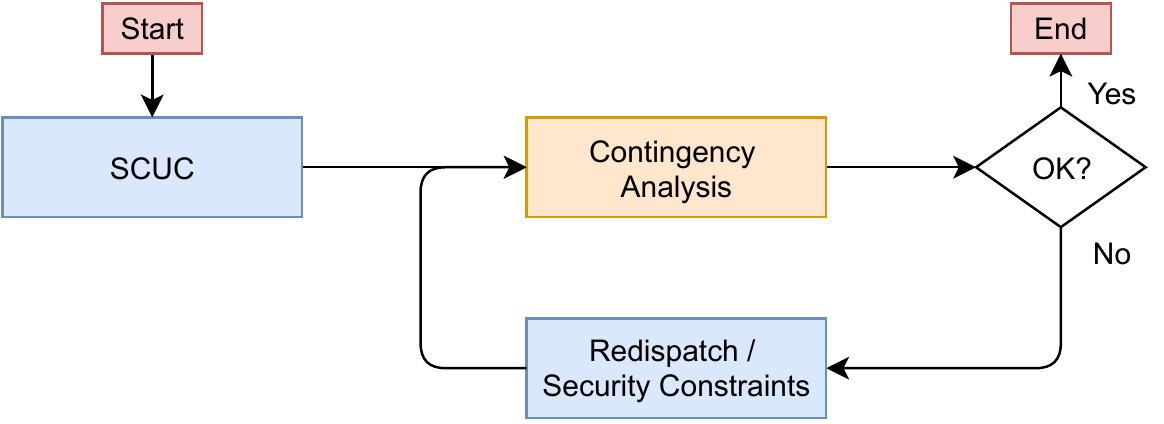}
    \caption{Flowchart of current practice SCUC with contingency analysis and heuristic corrections.}
    \label{fig:loop_current}
\end{figure}

{\color{black} 
In the current ISO practice, SCUC is solved  to determine the least-cost generator commitment and dispatch with respect to technical generator constraints, system constraints and security requirements, e.g., reserves.
The resulting schedules are then tested against a predefined set of contingencies to ensure that potential power mismatches or equipment overloads are within acceptable ranges or can be alleviated by the available resources.
If these security requirements can not be met, committed generators are re-dispatched using ad-hoc and out-of-optimization interventions (e.g., by means of constrained re-runs of parts of the SCUC \cite{al2014role}).
Figure~\ref{fig:loop_current} shows a schematic overview of this iterative approach.
Note that if no acceptable dispatch of the committed generators can be found, a re-run of the unit commitment with additional constraints may be necessary. 
}
Below we present the base SCUC and contingency analysis formulations, and discuss the required attributes for contingency reserves. 

\subsection{Base SCUC}
\label{sec:base_scuc}

The optimal generation commitments, power dispatch, and reserve allocations are determined by solving the following network- and security-constrained unit commitment optimization, \cite{al2014role}:
\allowdisplaybreaks
\begin{subequations}
\begin{align}
\min \quad 
    & \sum_{t\in\set{T}} \sum_{g\in\set{G}} t_{g,t} + u_{g,t} C_{g}^{0} + v_{g,t} C_{g}^{SU} + w_{g,t} C_{g}^{SD} \label{scuc:objective}\\
\text{s.t. }\forall t\in\set{T}: \hspace{-1cm}& \nonumber \\
    & t_{g,t} \ge p_{g,t} C_{1,o,g} + C_{0,o,g}, \quad \forall g\in\set{G}, \forall{o}\in\set{O} \label{scuc:pwlc}\\
    & u_{g,t} P^{\min}_g \le p_{g,t}, \quad \forall g\in\set{G} \label{scuc:gen_lower_limit} \\
    & u_{g,t} P^{\max}_g \geq p_{g,t} + r_{g,t}^S , \quad \forall g\in\set{G} \label{scuc:gen_upper_limit} \\
    & \sum_{s=t-UT_g-1}^t v_{g,s} \leq u_{g,t} , \quad \forall g\in\set{G} \label{scuc:uptime}\\
    & \sum_{s=t-DT_g-1}^t w_{g,s} \leq 1-u_{g,t} , \quad \forall g\in\set{G} \label{scuc:downtime}\\
    & v_{g,t} - w_{gt} = u_{g,t} - u_{g,t-1} , \quad \forall g\in\set{G} \label{scuc:su_sd_indicator}\\
    & p_{g,t} - p_{g,t-1} \leq R_g^{60} u_{g,t-1} + v_{g,t} {\color{black}P_{g}^{\min}} , \quad \forall g\in\set{G} \label{scuc:ramp_up}\\
    & p_{g,t-1} - p_{g,t} \leq R_g^{60} u_{g,t} + w_{g,t} {\color{black}P_{g}^{\min}} , \quad \forall g\in\set{G}\label{scuc:ramp_down}\\
    & \mathbb{1}_{w\not\in\set{W}^{\text{D}}} \overline{p}_{w,t} \leq  p_{w,t} \leq \overline{p}_{w,t}, \quad \forall w \in\set{W} \label{scuc:res_constraint} \\
    & f_{ij,t} = B_{ij}(\theta_{i,t} - \theta_{j,t}) , \quad  \forall ij\in\set{L} \label{scuc:power_flows} \\
    & \theta_{ref,t} = 0 \label{scuc:reference_bus} \\
    & -f^{\max}_{ij} \leq f_{ij,t} \leq f^{\max}_{ij} , \quad  \forall ij\in\set{L} \label{scuc:power_flow_limits}\\
    & \sum_{g\in\set{G}_i} p_{g,t} + \sum_{w\in\set{W}_i} p_{w,t} + \sum_{j:ij\in\set{L}}f_{ij,t} - \sum_{j:ji\in\set{L}}f_{ji,t} = D_{i,t}, \nonumber \\
        & \hspace{5cm} \forall i \in \set{N} \label{scuc:system_balance}\\
    & r_t^{tot} \leq \sum_{g\in\set{G}} (r_{g,t}^S + r_{g,t}^{NS}) \label{scuc:total_reserve} \\
    & r_t^{tot} \geq R^D \sum_{i\in\set{N}} D_{i,t} \label{scuc:min_reserve}\\
    & r_t^{tot} \geq p_{g,t} + r_{g,t}^S  , \quad \forall g\in\set{G}\label{scuc:max_reserve}\\
    & \sum_{g\in\set{G}}r_{g,t}^S \ge R^S r_t^{tot} \label{scuc:spinning_requirment} \\
    & {\color{black}r_{g,t}^{NS} \ge (1-u_{g,t})  P_{g}^{\min} , \quad \forall g\in\set{G}^{\text{FS}}} \label{scuc:nonspinning_min} \\
    & {\color{black}r_{g,t}^{NS} \le (1-u_{g,t})  P_{g}^{\max} , \quad \forall g\in\set{G}^{\text{FS}}}\label{scuc:nonspinning_max}\\
    & {\color{black}r_{g,t}^{NS} \le (1-u_{g,t}) R_{g}^{10} , \quad \forall g\in\set{G}^{\text{FS}}}\label{scuc:nonspinning_ramp}\\
    & {\color{black}r_{g,t}^{NS} = 0, \quad \forall g\in\set{G}^{\overline{\text{FS}}}} \label{scuc:nonspinning_nonfs} \\
    & r_{g,t}^{S} \le R_{g}^{10} , \quad \forall g\in\set{G} \label{scuc:spinning_ramp}\\
    & u_{g,t} \in \{0,1\} , \quad \forall g\in\set{G} \label{scuc:u_definiton}\\ 
    & 0 \le v_{g,t}, w_{g,t} \leq 1 , \quad \forall g\in\set{G}, \label{scuc:v_w_definiton}
\end{align}%
\label{mod:scuc}%
\end{subequations}%
\allowdisplaybreaks[0]%
where $\mathbb{1}_{x\in\set{X}}$ denotes the indicator function, which takes the value of $1$ if $x\in\set{X}$, and $0$ otherwise.
Objective \cref{scuc:objective} minimizes the system cost given by piecewise linear generator cost functions defined in \cref{scuc:pwlc}, no-load costs $C_g^0$, start-up costs $C_g^{SU}$ and shut-down costs $C_g^{SD}$.
Capacity limits of generators are enforced in \cref{scuc:gen_lower_limit,scuc:gen_upper_limit}.
Constraints \cref{scuc:uptime,scuc:downtime,scuc:su_sd_indicator} relate binary variables $u_{g,t}$, $v_{g,t}$ and $w_{g,t}$ that denote  commitment, start-up and shut-down decisions, respectively. 
Commitment changes are restricted by minimum up- and down-time limits enforced in \cref{scuc:uptime,scuc:downtime}. 
Note that it is sufficient to explicitly define $u_{g,t}$ as binary in \cref{scuc:u_definiton}, while  $v_{g,t}$ and $ w_{g,t}$ can be continuous within interval $[0,1]$ as in \cref{scuc:v_w_definiton}.
Constraints \cref{scuc:ramp_up,scuc:ramp_down} enforce generator ramping limits.\footnote{%
{\color{black}
Note that the formulation in this paper not model the precise the start-up and shut-down trajectories of the generators. 
For generators that require more than one time step (1 hour) to start-up/shut-down to/from $P_g^{\min}$, this inaccuracy must either be compensated by real-time system operations or by higher fidelity model formulations, e.g., as proposed in \cite{morales2012tight}.
Such models could be adopted in this paper without any methodical adjustments. However, we opt for the formulation in \cref{mod:scuc} as it is in line with some relevant publications, e.g.,  \cite{singhal2018data,pandvzic2013comparison,al2014role}, and allows for describing the proposed approach with reduced model complexity. 
}%
}
Renewable generation, e.g., grid-scale wind and solar power plants, is accounted for in set $\set{W}$. Constraint \cref{scuc:res_constraint} ensures that generation $p_{w,t}$ of renewable generator $w$ is lower than its forecast availability $\overline{p}_{w,t}$, if generator $w$ is dispatchable, i.e., $w\in\set{W}^D$, or equal to $\overline{p}_{w,t}$, if $w$ is not dispatchable, i.e., $w\not\in\set{W}^D$. 
The dc power flow equations, reference bus definition and thermal power flow limits are modeled as in \cref{scuc:power_flows,scuc:reference_bus,scuc:power_flow_limits}. 
Eq. \cref{scuc:system_balance} ensures the nodal power balance by accounting for the  generation/load injections and power flows at all nodes.
Finally, \cref{scuc:total_reserve,scuc:min_reserve,scuc:max_reserve,scuc:spinning_requirment,scuc:nonspinning_min,scuc:nonspinning_max,scuc:nonspinning_ramp,scuc:spinning_ramp} enforce contingency reserve requirements.
Specifically, reserve must cover at least the outage of the largest generator, \cref{scuc:max_reserve}, or a fraction $R^D$ of system demand, \cref{scuc:min_reserve}.  In turn, reserve consists of spinning and non-spinning portions, \cref{scuc:total_reserve}, whereas spinning reserve must be at least  $R^S$ times the total reserve.
Common values for $R^S$ and $R^D$  are $50\%$ and $7\%$, e.g., as in the current CAISO practice  \cite{al2014role}. However, other requirements are possible \cite{epri_reserve_dimensioning} to accommodate specific risk attitudes of the system operator.
In the model of \cref{mod:scuc}, only fast-start units (set $\set{G}^{\text{FS}}$) are allowed to provide non-spinning reserve, \cref{scuc:nonspinning_max,scuc:nonspinning_ramp,scuc:nonspinning_ramp,scuc:nonspinning_nonfs}, and all reserves are limited by the short-term (10-min) ramp-rate in \cref{scuc:nonspinning_ramp,scuc:spinning_ramp}.
{\color{black} 
Note that \cref{scuc:nonspinning_max,scuc:nonspinning_min,scuc:nonspinning_ramp} require fast-start units to reach their minimal production level within 10 minutes, which is in line with common eligibility criteria for non-spinning reserve providers \cite{isone2015fast}.
}

\subsection{Corrective Contingency Analysis}
\label{sec:corrective_contingency_analysis}

The commitment, dispatch and reserve decisions obtained from \cref{mod:scuc} are then evaluated for feasibility using a set of credible contingency scenarios. Thus, each contingency scenario is indexed as $c\in\set{C}$ and sets $\set{G}^c\subseteq\set{G}$ and $\set{L}^c\subseteq\set{L}$ contain indices of generators and lines that are unavailable during contingency $c$.
Next, the following optimization is solved for each contingency $c$:

\allowdisplaybreaks
\begin{subequations}
\begin{align}
    \min \quad \hspace{-0.5cm}&\hspace{0.5cm} \sum_{t\in\set{T}} [ C^{\text{LS}} \sum_{i\in\set{N}} s_{i,t,c}^{\text{LS}} + C^{\text{overl}} \sum_{ij\in\set{L}}(s_{ij,t,c}^+ + s_{ij,t,c}^-)] \label{corrective:objective}\\
\text{s.t. }\forall t\in\set{T}: \hspace{-1cm} & \nonumber \\
    & f_{ij,t,c} = B_{ij}(\theta_{i,t,c} - \theta_{j,t,c}) , \quad \forall ij\in\set{L}\setminus\set{L}^c \label{corrective:power_flows} \\
    & \theta_{ref,t} = 0 \label{corrective:slack}\\
    & -\!\mathbb{1}_{ij\not\in\set{L}^c}(f^{E}_{ij} + s_{ij,t}^-) \leq f_{ij,t,c} \leq \mathbb{1}_{ij\not\in\set{L}^c}(f^{E}_{ij} + s_{ij,t}^+), \nonumber \\ 
        & \hspace{55mm}  \forall ij\in\set{L} \label{corrective:line_limit_emergency} \\
    & \mathbb{1}_{w\not\in\set{W}^{\text{D}}} \overline{p}_{w,t} \leq  p_{w,t,c} \leq \overline{p}_{w,t}, \quad \forall w \in\set{W} \label{corrective:res_constraint} \\
    & \sum_{g\in\set{G}_i} p_{g,t,c} \!+\!\!\!\sum_{w\in\set{W}_i} p_{w,t,c} \!+\!\!\! \sum_{j:ij\in\set{L}}f_{ij,t,c} \!-\!\!\!\sum_{j:ji\in\set{L}}f_{ji,t,c}  \nonumber \\
        & \hspace{35mm} \!=\! D_{i,t} - s_{i,t,c}^{\text{LS}},\ \forall i \in \set{N} \label{corrective:power_balance}\\
    & s^{\text{LS}}_{i,t} \le \max\{0, D_{i,t}\}, \quad\forall i \in \set{N} \label{corrective:ls_limit} \\
    & u^*_{g,t} \!\mathbb{1}_{g\not\in \set{G}^c} (p^*_{g,t} \!\!-\! R_g^{10}) \!\leq p_{g,t,c} \!\leq u^*_{g,t} \!\mathbb{1}_{g\not\in \set{G}^c} (p^*_{g,t} \!\!+\! R_g^{10}), \nonumber\\
        & \hspace{55mm} \forall g \in \set{G} \label{corrective:ramping_constraint} \\
    & u^*_{g,t} \mathbb{1}_{g\not\in \set{G}^c} P_g^{\min} \leq p_{g,t,c} \leq u^*_{g,t} \mathbb{1}_{g\not\in \set{G}^c} P_g^{\max}, \nonumber \\
        & \hspace{55mm} \forall g \in \set{G}^{\overline{\text{FS}}} \label{corrective:non_fs_limits} \\
    & 0 \leq p_{g,t,c} \leq \mathbb{1}_{g\not\in \set{G}^c} P_g^{\max} , \quad \forall g \in \set{G}^{\text{FS}}  \label{corrective:fs_limits}\\
    & s_{i,t,c}^{\text{LS}} \ge  0 , \quad \forall i \in \set{N} \\
    & s_{ij,t,c}^+, s_{ij,t,c}^- \ge 0 , \quad \forall ij \in \set{L},
\end{align}%
\label{mod:corrective}%
\end{subequations}%
\allowdisplaybreaks[0]%
where $p^*_{g,t}$ and $u^*_{g,t}$ are  pre-contingency decisions, i.e., the results obtained directly from solving \cref{mod:scuc}, and set $\set{G}^{\overline{\text{FS}}}=\set{G}\setminus\set{G}^{\text{FS}}$ is the set of generators that cannot be synchronized in real time, e.g., within less than 10 minutes. 
For every time period $t$, \cref{mod:corrective} re-dispatches available generators so that load shedding $s_{i,t,c}^{\text{LS}}$ and positive and negative transmission line overloads $s_{ij,t,c}^+$, $s_{ij,t,c}^-$ are minimized as given by the  objective function in \cref{corrective:objective}.
Constraints \cref{corrective:power_flows,corrective:slack,corrective:line_limit_emergency} enforce power flow equations and limits for all available  lines, i.e.,  $\forall ij\in\set{L}\setminus\set{L}^c$.
In addition to potential line overloads $s_{ij,t,c}^+$ and $s_{ij,t,c}^-$, the maximum thermal capacity of each line is set to its emergency rating $f_{ij}^{E}\geq f_{ij}^{\max}$, which can be maintained for a short period of time. 
{\color{black}
Note that $f_{ij}^{E} = f_{ij}^{\max}$ is possible, e.g., when power flows are limited to comply with voltage stability criteria.}
Nodal power balance is enforced in \cref{corrective:power_balance} with a possibility of  load shedding $s_{i,t,c}^{\text{LS}}$. Load shedding at each time $t$ and bus $i$ is limited by the nodal load $D_{i,t}$ as enforced in \cref{corrective:ls_limit}.
Constraints \cref{corrective:ramping_constraint,corrective:non_fs_limits,corrective:fs_limits} restrict the contingency dispatch by the short-term ramping  and output limits of each generator.

\subsection{Deliverability of Reserves}
Reserves allocated in \cref{mod:scuc} may not be deliverable under specific contingency scenarios analyzed using the model in \cref{mod:corrective} due to generation limits and network congestion. Such instances are defined for contingency $c$ when \cref{mod:corrective} returns  $\sum_{t\in\set{T}} \sum_{i\in\set{N}} s_{i,t}^{\text{LS}} +  \sum_{t\in\set{T}} \sum_{ij\in\set{L}}(s_{ij,t,c}^+ + s_{ij,t,c}^-) > 0$, i.e., the loss of generators $\set{G}^c$ and/or lines $\set{L}^c$ can not be corrected by the available reserves without overloading transmission lines  or load shedding. 
Alternatively, the base SCUC  formulation in \cref{mod:scuc} can be modified to endogenously ensure the deliverability of reserves for every $c\in\set{C}$. 
Formally, such a modification can be written as
\begin{subequations}
\begin{align}
    \min \quad 
    & \sum_{t\in\set{T}} \sum_{g\in\set{G}} t_{g,t} + u_{g,t} C_{g}^{0} + v_{g,t} C_{g}^{SU} + w_{g,t} C_{g}^{SD} \label{res_scuc:objective}\\
    \text{s.t. }\forall t\in\set{T}: \hspace{-1cm}& \nonumber \\
    & \text{\cref{scuc:pwlc}--\cref{scuc:v_w_definiton}} \nonumber \\
    & \Delta f_{ij,t,c} \leq f_{ij}^{E} - f_{ij,t}, \quad \forall ij\in\set{L},\ \forall c\in\set{C} \label{res_scuc:genout_delatf_upper}\\
    & -\Delta f_{ij,t,c} \leq f_{ij}^{E} + f_{ij,t}, \quad \forall ij\in\set{L},\ \forall c\in\set{C} \label{res_scuc:genout_delatf_lower},
\end{align}%
\label{mod:res_scuc}%
\end{subequations}%
where \cref{res_scuc:genout_delatf_upper,res_scuc:genout_delatf_lower} ensure that changes in power flow $\Delta f_{ij,t,c}$ due to contingency $c$ can be accommodated by the system.
However, the implicit computation of $\Delta f_{ij,t,c},\ \forall c\in\set{C}$ and enforcing  \cref{res_scuc:genout_delatf_upper,res_scuc:genout_delatf_lower} $\forall c \in\set{C}$ generally leads to computationally intractable SCUC problems, even for small networks \cite{capitanescu2011state,madani2016constraint}.
To overcome this, we propose below a computationally tractable and risk-aware approach to \cref{mod:res_scuc} by iteratively learning a linear relationship between $\Delta f_{ij,t,c}$ and scheduled reserves $r_{g,t}$.
{\color{black} 
Notably, if $\Delta f_{ij,t,c}$ can be estimated accurately, then the SCUC modification in \cref{mod:res_scuc} can ensure risk-aware reserve delivery by adding no additional variables and two linear constraints per considered contingency.}
{\color{black}
The similarity with established SCUC formulations makes the proposed approach suitable for gradual implementation and adoption of risk-management methods in real-world systems.}

\section{Risk-Aware Reserve Allocation}

\begin{figure}
    \centering
    \includegraphics[width=0.95\linewidth]{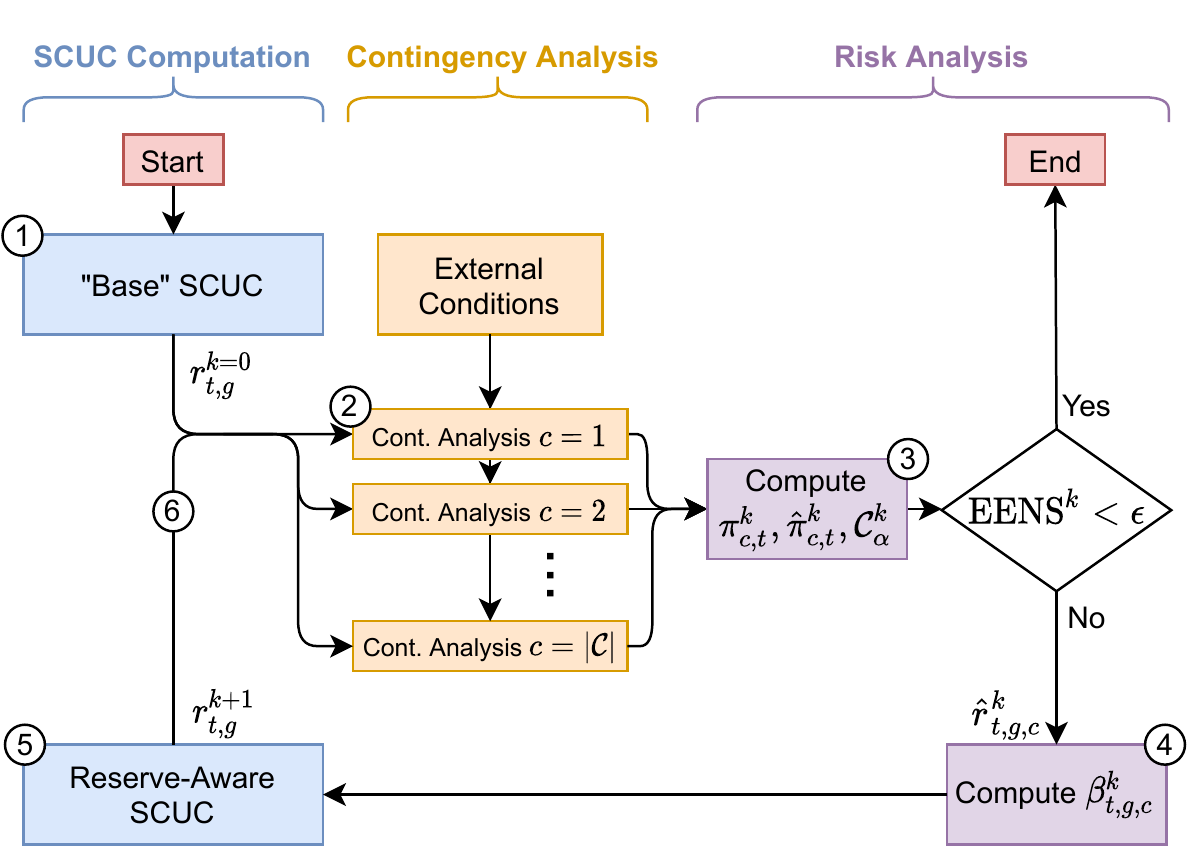}
    \caption{Flow diagram for activation factors training}
    \label{fig:loop_illustration}
\end{figure}

Current SCUC methods used by real-world ISOs ignore the probability of individual contingency scenarios, which leads to commitment, dispatch and reserve decisions that misestimate  risk exposure of the system and does not reflect actual system risk levels \cite{ortega2007optimizing}.
The risk associated with a contingency scenario can be estimated as  the probability of this scenario times its severity. Notably, severity must be considered from the system perspective, i.e., it must capture the ability of the power system to remain in a stable operational state and continue serving the system load. 
For example, if it can be guaranteed that the loss of a generator can be safely compensated by available reserves, then this contingency poses no risk to system operation. 
A suitable contingency risk metric is the \textit{expected energy not served} (EENS), i.e., the amount of unserved load after a contingency multiplied by the probability of that contingency. 
However, EENS computation requires evaluating possible post-contingency system states. Solving a one-shot SCUC with an internalized EENS-based risk evaluation is computationally demanding and requires a significant modification to the original SCUC formulation (e.g., see \cite{fernandez2016probabilistic}), which is undesirable from the viewpoint of real-world ISOs as it reduces transparency, accountability and trustworthiness of the resulting decisions. 

To enable a risk-aware reserve allocation process, it is critical  to (i)  ensure the reserve deliverability  with a high probability and (ii)  trade off the risk-adjusted cost and benefits of reserve procurement with  minimal  alterations to base SCUC practice, which has earned the trust of market participants. To this end, the iterative approach in  Fig.~\ref{fig:loop_illustration} enhances the  current practice to include risk adjustments accounting for different likelihoods of  contingency scenarios.

Each iteration of the algorithm in  Fig.~\ref{fig:loop_illustration} performs SCUC computation, contingency analysis and a risk analysis.  
The detailed process is itemized below where each step corresponds to the circled numbers in Fig.~\ref{fig:loop_illustration}:
\begin{enumerate}
    \item Solve the base SCUC formulation in \cref{mod:scuc} with given reserve requirements.
    \item 
    {\color{black}
     Perform contingency analyses as in \cref{mod:corrective} for the  predefined set  of credible contingencies ($\set{C}$), where each contingency scenario $c\in\set{C}$ considers either a single or a combined outage of one or multiple generation or transmission assets}. 
    Additionally, this step considers different external disturbances such as extreme weather events and VRES fluctuations are considered.
    \item  
    {\color{black}
    Calculate the risk of all contingencies using a suitable risk metric (e.g., EENS) based on their respective contingency probability and impact (captured by $\pi_{c,t}$, $\hat{\pi}_{c,t}$, $\set{C}_\alpha$, see Section~\ref{ssec:eens_calculation} below for details).
    }
    If the resulting risk level is below a given threshold $\epsilon$, the process stops.
    \item If the risk level is above $\epsilon$, update the reserve activation factors and compute worst-case contingency probabilities as described in Sections~\ref{ssec:reserve_activation_factors} and \ref{ssec:risk_adjusted_reserve_deliverarbility} below. 
    \item Re-run the reserve-aware modification of \cref{mod:scuc} (to which we refer as ResA-SCUC) by additionally enforcing post-contingency power flow constraints as described in  Section~\ref{ssec:reserve_activation_factors} below.
    \item Repeat until a desired level of risk given by $\epsilon$ is achieved. 
\end{enumerate}

The following subsections describe the formulation and computations needed to accommodate the procedure in Fig.~\ref{fig:loop_illustration} within the current practice. To this end, we describe  necessary, rather non-intrusive modifications to the base SCUC formulation in \cref{mod:scuc} to obtain the ResA-SCUC used in step (5) of the procedure in Fig.~\ref{fig:loop_illustration}.

\subsection{Reserve Activation Factors}
\label{ssec:reserve_activation_factors}

To internalize the effects of post-contingency reserve activation into the base SCUC formulate as in \cref{mod:res_scuc}, we use \textit{reserve activation factors} (RAFs) denoted as  $\beta_{g,t,c}$.
RAFs establish a functional connection between the  scheduled reserves $r_{g,t} = r_{g,t}^{\text{S}}+r_{g,t}^{\text{NS}}$ and corrective generation actions $\hat{r}_{g,t,c} = p_{g,t,c} - p^*_{g,t}$.
First, we consider reserve as a strictly positive corrective measure to respond to a credible generation contingency.
Similarly to \cite{singhal2018data}, we assume that the relationship between $r_{g,t}$ and $\hat{r}_{g,t,c}$ can be approximated by the  following linear function:
\begin{align}
    \hat{r}_{g,t,c} \approx f(\beta_{g,t,c}, r_{g,t}) = \beta_{g,t,c} r_{g,t}, \quad \forall c \in \set{C}^{\text{G}},
    \label{eq:sensitivity_function}
\end{align}
where $\set{C}^{\text{G}}$ is the set of all contingency scenarios with generator outages. Thus, if parameters $\beta_{g,t,c}$ in \cref{eq:sensitivity_function} can be estimated accurately, they can be used to assure reserve deliverability for credible contingencies in the base SCUC formulation in \cref{mod:scuc} by adding the following set of constraints:
\allowdisplaybreaks
\begin{subequations}
\begin{align}
    & \Delta f_{ij,t,c} = \sum_{k\in\set{N}} \ptdf_{(ij)k}\Big(\sum_{g\in\set{G}_k} \beta_{g,t,c} r_{g,t} - \sum_{g\in\set{G}_k\cap\set{G}^c}p_{g,t}\Big), \nonumber \\
      &  \hspace{55mm} \forall ij \in \set{L},\ \forall c\in\set{C}^{\text{G}} \label{eq:delta_f_gen_outages}\\
    & \Delta f_{ij,t,c} \leq f_{ij}^{E} - f_{ij,t} \label{eq:genout_delatf_upper}\\
    & -\Delta f_{ij,t,c} \leq f_{ij}^{E} + f_{ij,t} \label{eq:genout_delatf_lower},
\end{align}%
\label{eq:raf_gen_constraints}%
\end{subequations}%
\allowdisplaybreaks[0]%
where $\ptdf_{(ij)k}$ is the power transfer distribution factor of nodal power injections at bus $k$ towards the active power flow in transmission line $ij$.

{\color{black} 
In contrast to reserve activation during generation outages, which usually only requires the activation of upward reserves \cite{singhal2018data}, re-dispatch actions following credible transmission line contingencies often require both upward and downward flexibility of generators. 
}
The amount of this flexibility available for reserve activation depends on the commitment and  dispatch capacity of generators, and therefore we compute it similarly to \cref{eq:sensitivity_function} using the following  relationship:
\begin{align}
    \hat{r}_{g,t,c} \approx f(\beta_{g,t,c}, R_g^{10}) = \beta_{g,t,c} R_g^{10}, \quad \forall c \in \set{C}^{\text{L}} \label{eq:ramping_sensitivity}
\end{align}
where $\set{C}^L$ is the set of contingency scenarios with transmission line contingencies. \footnote{
{\color{black} 
 In some relatively rare cases, the downward re-dispatch of a generator may also be required to deal with  generator outage, e.g., to free up transmission capacity. While we have not observed such behavior in our numerical experiments, which is in line with related research \cite{singhal2018data}, the approach described in Eqs.~\cref{eq:ramping_sensitivity,eq:delta_f_line_outages,eq:factor_computation_up,eq:factor_computation_down} can also be used for generation contingencies that require downward flexibility.} 
}
{\color{black} 
Note that the total available downward reserve of generator $g$ may be lower than $R_g^{10}$, if $p_{g,t}^* - P_g^{\min}\le R_g^{10}$. In this case the absolute range of $\beta_{g,t,c}$ would be smaller than one. 
}

To inform the base SCUC formulation of both upward and downward flexibility available during transmission line contingencies, we distinguish between positive and negative RAFs such that $\beta_{g,t,c}^+\in[0,1]$ or  $\beta_{g,t,c}^-\in[-1,0]$ model the expected flexibility  under the condition that the activation is positive or negative.  
Furthermore, to internalize the impact of transmission line contingencies, we must also consider the redistribution of line flows due to changes in network topology. 
Therefore, $\Delta f_{ij,t,c}$ must be updated for each contingency scenario as follows:
\allowdisplaybreaks
\begin{subequations}
\begin{align}
\begin{split}
    \Delta f_{ij,t,c}^+ =&
    \sum_{k\in\set{N}} F_{(ij)k}^c(\sum_{g\in\set{G}_k} p_{g,t} + \sum_{w\in\set{W}_k} p_{w,t}) \\
     &\hspace{-15mm} + \max\Big\{0,\sum_{k\in\set{N}}\!(\ptdf_{(ij)k}\! + F^{\color{black}c}_{(ij)k})\sum_{g\in\set{G}_k} \beta_{g,t,c}^+ u_{g,t} R_g^{10}, \\
    &\hspace{-18mm} \qquad \sum_{k\in\set{N}}\!(\ptdf_{(ij)k}\! + F^{\color{black}c}_{(ij)k})\sum_{g\in\set{G}_k} \beta_{g,t,c}^- u_{g,t} R_g^{10}\Big\},\ \forall c\in\set{C}^{\text{L}}
\end{split}\\
\begin{split}
    \Delta f_{ij,t,c}^- =&
    \sum_{k\in\set{N}} F^c_{(ij)k}(\sum_{g\in\set{G}_k} p_{g,t} + \sum_{w\in\set{W}_k} p_{w,t}) \\
    &\hspace{-15mm} + \min\Big\{0,\sum_{k\in\set{N}}(\ptdf_{(ij)k} + F^{\color{black}c}_{(ij)k})\sum_{g\in\set{G}_k}  \beta_{g,t,c}^+ u_{g,t} R_g^{10},\\
    &\hspace{-18mm} \qquad \sum_{k\in\set{N}}(\ptdf_{(ij)k} + F^{\color{black}c}_{(ij)k})\sum_{g\in\set{G}_k}  \beta_{g,t,c}^- u_{g,t} R_g^{10}\Big\},\ \forall c\in\set{C}^{\text{L}}
\end{split}\\
   \Delta f_{ij,t,c}^+ \leq& f_{ij}^{E} - f_{ij,t} \label{eq:lineout_delatf_upper}\\
     -\Delta f_{ij,t,c}^- \leq& f_{ij}^{E} + f_{ij,t}, \label{eq:lineout_delatf_lower}
\end{align}%
\label{eq:delta_f_line_outages}%
\end{subequations}%
\allowdisplaybreaks[0]%
where $F^c_{(ij)k}$ captures the sensitivity of the power flow \textit{change} on line $ij$ to power injections at bus $k$ during contingency $c$. 
Sensitivity $F^c_{(ij)k}$ is given by $F^c_{(ij)k} = \lodf_{(ij)\set{L}^c}\ptdf_{\set{L}^c k}$. 
The entries of $(1\times|\set{L}^c|)$-vector $\lodf_{(ij)\set{L}^c}$ are the \textit{load outage distribution factors} of tripped lines $\set{L}^c$ towards line $ij$. The entries of $(|\set{L}^c\times 1|)$-vector $\ptdf_{\set{L}^c k}$ are the PTDFs of bus $k$ towards lines $\set{L}^c$.
See, e.g., \cite{weinhold2020fast}.
{\color{black} 
The resulting sensitivities $(\ptdf_{(ij)k} + F^c_{(ij)k})$ are are also called \textit{outage transfer distribution factors}.
}

We note that if a  contingency scenario considers both transmission and generation outages, then both Eqs. \cref{eq:raf_gen_constraints} and \cref{eq:delta_f_line_outages} must be added to the base SCUC formulation,
{\color{black} 
which will ensure that each transmission line can withstand the largest estimate of $\Delta f_{ij,t,c}, \Delta f_{ij,t,c}^+, \Delta f_{ij,t,c}^-$}.
{\color{black} 
Further, we note that although the explicit separation of  $\Delta f_{ij,t,c}^+$ and $\Delta f_{ij,t,c}^-$ in \cref{eq:delta_f_line_outages} provides an additional degree of fidelity to the model, which allows for accommodating more general combinations of $\beta_{g,t,c}^+ $ and $\beta_{g,t,c}^-$, a reduction to a single variable ($\Delta f_{ij,t,c}$) is possible.
}
Introducing \cref{eq:raf_gen_constraints} and/or \cref{eq:delta_f_line_outages} to the base SCUC formulation in  \cref{mod:scuc} makes it possible to adjust reserve deliverability under different credible contingencies and adjust the risk of these outages. As shown in Fig.~\ref{fig:loop_illustration}, this modification of the base SCUC is the reserve-aware SCUC (ResA-SCUC).

\subsection{Learning Reserve Activation Factors}

For every iteration $k$ of the procedure proposed in Fig.~\ref{fig:loop_illustration}, RAFs $\beta_{g,t,c}^k$ for credible generator contingencies can be obtained from the following learning process:
\begin{align}
    \beta_{g,t,c}^k = \max\Big\{\frac{\hat{r}_{g,t,c}^{k-1}}{\color{black}r_{g,t}}, \lambda \frac{\hat{r}_{g,t,c}^{k-1}}{\color{black}r_{g,t}} + (1-\lambda)\beta_{g,t,c}^{k-1}\Big\},
    \label{eq:factor_computation}
\end{align}
where $\beta_{g,t,c}^{0} = 0$ and parameter $\lambda\in[0,1]$ defines the ``memory decay'' of the process.
If $\lambda = 1$, all factors $\beta_{g,t,c}^{k}$ depend only on the SCUC and contingency analyses results of the  previous iteration indexed as  $k-1$. On the other hand, if $\lambda = 0$, reserve activation from all previous iterations are considered, i.e, $\beta_{g,t,c}^{k} \geq \beta_{g,t,c}^{k-1} \geq ... \geq \beta_{g,t,c}^{0}$, where  $\beta_{g,t,c}^{k}\in[0,1]$. 

\begin{remark}
\label{rem:memory_decay_rate}
In our experiments a relatively  low value of the  memory decay rate, e.g., $\lambda=0$, is the most effective. 
This is because if a certain $r_{g,t}$ can be scheduled cheaply by the SCUC, but can never be fully or partially delivered during contingency analysis, this knowledge should be kept through all iterations. As a result, the SCUC must schedule more expensive, but deliverable reserves. 
\end{remark}

\begin{remark}
The learning process in  \cref{eq:factor_computation} is also more effective than regression-based approaches, such as in \cite{singhal2018data}, because many RAFs $\beta_{g,t,c}$ tend to be either $1$ or to $0$, see Section~\ref{ssec:reserve_activation_factors_case_study} below. 
Therefore, for generators that have consistent reserve activation factors $\beta_{g,t,c}=1$ across iterations, regression approaches would lead to a slope of $0$ and an intercept of $1$.
Forcing the intercept to $0$, however, may bias the regression and misestimate the RAF for reserves that are only partially activated.
\end{remark}

Since we consider both  positive and  negative reserve activation RAFs for credible line contingencies in  \cref{eq:factor_computation}, we also differentiate at each iteration:
\allowdisplaybreaks
\begin{align}
    \beta_{g,t,c}^{+,k} = \begin{cases}
    \max\Big\{\frac{\hat{r}_{g,t,c}^{k-1}}{R^{10}_{g}}, \lambda \frac{\hat{r}_{g,t,c}^{k-1}}{R^{10}_{g}} + (1-\lambda)\beta_{g,t,c}^{+,k}\Big\} & \text{if }  \hat{r}_{g,t,c} \ge 0\\
    \beta_{g,t,c}^{+,k-1} & \text{else.}
    \end{cases} \label{eq:factor_computation_up} \\
    \beta_{g,t,c}^{-,k} = \begin{cases}
    \min\Big\{\frac{\hat{r}_{g,t,c}^{k-1}}{R^{10}_{g}}, \lambda \frac{\hat{r}_{g,t,c}^{k-1}}{R^{10}_{g}} + (1-\lambda)\beta_{g,t,c}^{-,k}\Big\} & \text{if }  \hat{r}_{g,t,c} \le 0\\
    \beta_{g,t,c}^{-,k-1} & \text{else.}
    \end{cases}\label{eq:factor_computation_down}
\end{align}
\allowdisplaybreaks[0]

{\color{black} 
\subsection{Convergence}

There are three noteworthy remarks on the convergence of the proposed approach, eventually reaching the stopping criterion ($\eens^k < \epsilon$) as shown in Fig.~\ref{fig:loop_illustration}.
First, it is theoretically possible that $\eens^k < \epsilon$ can not be achieved for the given system, i.e., it is possible that certain contingencies cause unavoidable load shedding independent of the procured reserves. This may occur, for example, when one or multiple buses become separated from the remaining system. However, such occurrences typically point to underlying design, rather than operational risk. As such,  real-world systems are usually designed to avoid such events, and/or their likelihood is sufficiently low, and/or such events extend beyond the notion of credible contingencies that must be preventively dealt with. 
{\color{black} 
Hence, if the proposed approach fails to reach the desired EENS value, the ISO may have to employ additional relaxations such as load reduction or transmission demand curves \cite{nyiso2020manual12}. 
}

Second, because the proposed approach relies on solving a unit commitment problem, which is non-convex, the EENS might not decrease monotonically (see a detailed discussion in \cite{ortega2007optimizing}). We have observed this effect, for example, when the system must choose between committing generators with similar (or identical) parameters. If only one generator from a set of  units with comparable parameters is needed and this unit is, after iteration $k$, associated with a non-zero reserve activation factor, the model may not choose this unit again in iteration $k+1$ to avoid reserving transmission capacity for the expected reserve activation associated with this unit. As a result, EENS may switch and then oscillate between a higher and a lower value. While this effect can be overcome by penalizing commitment changes after a certain number of iterations, we have not observed it in the experiments shown in this paper. 
Finally, convergence speed depends on the choice of  decay rate $\lambda$. As outlined in Remark~\ref{rem:memory_decay_rate}, a lower value of rate $\lambda$ will consider all activated reserves from previous iterations and, thus, quickly lead to a robust solution that can achieve $\eens^k < \epsilon$. 
A larger $\lambda$, on the other hand, can lead to a less conservative solution but may increase the number of necessary iterations. 
As mentioned in Remark~\ref{rem:memory_decay_rate}, we found in our experiments that the additonal conservatism incurred by adopting a low value of  $\lambda$ is negligible.}

\subsection{Risk-Adjusted Reserve Deliverability}
\label{ssec:risk_adjusted_reserve_deliverarbility}

The iterative approach shown in Fig.~\ref{fig:loop_illustration} uses EENS as a stopping criterion and is therefore risk-aware. 
However, for all $c\in\set{C}$ and $\beta_{g,t,c}\neq 0$, the ResA-SCUC enforces post-contingency flow changes $\Delta f_{ij,t,c}$ in a \textit{robust} manner.
That is, constraints \cref{eq:genout_delatf_upper,eq:genout_delatf_lower,eq:lineout_delatf_upper,eq:lineout_delatf_lower}
can be written as:
\allowdisplaybreaks
\begin{align}
    \max_{c\in\set{C}}{\Delta f_{ij,t,c}} &\leq f_{ij}^E - f_{ij,t}
    \label{eq:robust_deltaf} \\
    \max_{c\in\set{C}}{-\Delta f_{ij,t,c}} &\leq f_{ij}^E + f_{ij,t} \\
    \max_{c\in\set{C}}\Delta f_{ij,t,c}^+ &\leq f_{ij}^{E} - f_{ij,t} \\
    \max_{c\in\set{C}} -\Delta f_{ij,t,c}^- &\leq f_{ij}^{E} + f_{ij,t}, 
\end{align}%
\allowdisplaybreaks[0]%
respectively.
As a result, each line $ij\in\set{L}$ maintains a security margin to sustain the worst-case flow change.

Instead of enforcing post-contingency flows in a robust manner as in \cref{eq:robust_deltaf}, contingency probabilities  $\pi_c^k,\ \forall c\in\set{C}$ can be used for risk-aware decision making and for  selecting a set of worst-case contingencies $\set{C}^{k}_\alpha$. Note that contingency probabilities $\pi_c^k$ may change in between iterations because they depend on the generator commitment status. See Sections~\ref{ssec:generator_reliability} and \ref{ssec:contingency_probabilites} below.
The  process of computing  $\set{C}^{k}_\alpha$ is shown in Algorithm~\ref{alg:risk_aware_cont_set}.

 \begin{algorithm}[t]
    \SetAlgoLined
    \SetKwInOut{Input}{input}\SetKwInOut{Output}{output}
    \Input{contingency risks $\eens_c^k,\ \forall c\in\set{C}$, contingency probabilities $\pi_c^k,\ \forall c\in\set{C}$, risk-level $\alpha$}
    \Output{set of worst-case contingencies $\set{C}^{k}_\alpha$, \\ 
    adjusted worst case probabilities $\hat{\pi}_c^{\alpha,k},\ \forall c\in\set{C}$}
    \Begin{
      $\set{C}^{k}_\alpha \leftarrow \emptyset$\;
      Sort $\{\eens_c^k,\ \forall c\in\set{C}\}$ and collect resulting contingency indices in set $\set{E} = \{e_1, e_2, \hdots, e_{|\set{C}|}\}$ such that $\eens_{e_1}^k\geq\eens_{e_2}^k\geq,\hdots,\eens_{e_{|\set{C}}|}^k$\;
      $i \leftarrow 1$\;
      \While{($\sum_{c\in\set{C}^{k}_\alpha}\pi_c^k\leq\alpha) \land (i\le|\set{C}|)$}{
        $\set{C}^{k}_\alpha \leftarrow \set{C}^{k}_\alpha \cup e_i$\;
        $i\leftarrow i+1$\;
      }
      \For{$c\in\set{C}$}{
        \eIf{$c\in\set{C}^{k}_\alpha$}{
            $\hat{\pi}_c^{\alpha,k} \leftarrow {\pi_c^k}/{\sum_{c\in\set{C}^{k}_\alpha}\pi_c^k}$
        }{
            $\hat{\pi}_c^{\alpha,k} \leftarrow  0$
        }
      }
      \KwRet{$\set{C}^{k}_\alpha$, $\{\hat{\pi}_c^{\alpha,k},\ \forall c\in\set{C}\}$}
     }
    \caption{Worst Case Contingencies at Iteration $k$}
    \label{alg:risk_aware_cont_set}
\end{algorithm}

For each contingency, Algorithm~\ref{alg:risk_aware_cont_set} also returns an adjusted conditional probability $\hat{\pi}_c^{\alpha,k}$ that captures the probability of contingency $c$ under the condition that one of the worst-case contingencies occurs. As a result, $\hat{\pi}_c^{\alpha,k}=0,\ \forall c\in\set{C}\setminus\set{C}^{k}_\alpha$.
Now, instead of enforcing \cref{eq:genout_delatf_upper} and/or \cref{eq:lineout_delatf_upper,eq:lineout_delatf_lower} for all $c\in\set{C}$, we can ensure feasibility of the $\alpha$-worst case expected power flows. As a result, \cref{eq:robust_deltaf} can be substituted with:
\begin{equation}
   \sum_{c\in\set{C}}\Delta f_{ij,t,c}\hat{\pi}_c^{\alpha,k} \leq f_{ij}^E - f_{ij,t}. \label{eq:riskadjusted_deltaf}
\end{equation}
Again, constraints \cref{eq:genout_delatf_lower,eq:lineout_delatf_upper,eq:lineout_delatf_lower} can brought in a risk-aware form analogously.
Note that this approach effectively recovers the $\alpha$-conditional value-at-risk ($\cvar_{\alpha}$) as defined in  \cite{rockafellar2007coherent,roveto2020co}.
As a result, if $\alpha=1$ the system is immunized against the expected change in power flows, i.e.,
\begin{equation}
 \sum_{c\in\set{C}}\Delta f_{ij,t,c}\hat{\pi}_c^{\alpha=1,k} = \sum_{c\in\set{C}}\Delta f_{ij,t,c}\pi_c^k,   
\end{equation}
and if $\alpha=0$ the system will be immunized against the single worst-case post-contingency power flow, i.e., 
\begin{equation}
    \sum_{c\in\set{C}}\Delta f_{ij,t,c}\hat{\pi}_c^{\alpha=0,k} = \max_{c\in\set{C}}{\Delta f_{ij,t,c}}.
\end{equation}

\section{Probability of Contingencies}
\label{sec:probability_of_contingenies}

To compute the EENS in step 3) of the loop shown in Fig.~\ref{fig:loop_illustration} and the risk-adjusted post-contingency flows as outlined in Section~\ref{ssec:risk_adjusted_reserve_deliverarbility}, we need to compute  probability $\pi_c$ of each contingency $c\in\set{C}$. 
Below we outline the underlying generator reliability model to obtain these probabilities accounting for both adverse conditions and failure to synchronize and discuss the computation of $\pi_c$ and EENS. 

\subsection{Generator Reliability}
\label{ssec:generator_reliability}

A typical reliability metric is the forced outage rate (FOR), i.e., the relative frequency of a generator not being available when it should have been \cite{allan2013reliability,ortega2008optimising}. 
Assuming that FOR is independent for every hour and that repair times are longer than the mission time $T_g$, i.e. 1 to 24 hours, it has been shown that the reliability of generator $g$ can be modeled via its \textit{outage replacement rate} (ORR):
\begin{equation}
   \orr_g(T_g) = \int_0^{T_g} q_g e^{-q_g t}d t = 1 - e^{-q_g T_g},
   \label{eq:orr}
\end{equation}
where $\orr_g(T_g)$ is the probability of the unplanned unavailability of generator $g$ during mission time $T_g$ with $q_g$ denoting the  FOR of that generator.
In line with  reliability models of generators from previous studies, e.g., \cite{ortega2007optimizing,fernandez2016probabilistic}, we set $T_g=1$ and obtain:  
\begin{equation}
    \orr_g(T_g) = \orr_g = 1 - e^{-q_g}.
\end{equation}

Additionally, whenever a generator is starting up, i.e., switching between the on and off states, it  may fail to synchronize. 
As shown in \cite{ortega2008optimising}, the reliability model in \cref{eq:orr} can be extended to consider this failure to synchronize as follows:
\begin{align}
    A_g &= (1-q_g^s)(1-\orr_g) \\
    U_g &= 1 - (1-q_g^s)(1-\orr_g),
\end{align}
where $A_g$ denotes the probability that the generator successfully synchronizes and is available, and $U_g$ denotes the probability that the generator is unavailable due to either failed synchronization or an unexpected outage. Parameter $q_g^s$ denotes the relative frequency of synchronization failures and can be estimated from historical data \cite{ortega2008optimising}.
Using the scheduling results from \cref{mod:scuc}, we can compute the probability of each generator $g$ to be unavailable at time $t$ as:
\begin{align}
    U_{g,t} = u_{g,t}^*(1 - (1-v_{g,t}^*q_g^s)(1-\orr_g)),
    \label{eq:generator_unavailability}
\end{align}
where $u_{g,t}^*\in\{0,1\}$ and $v_{g,t}^*\in\{0,1\}$ are the commitment and start-up decisions produced by the base SCUC model in \cref{mod:scuc} and described in Section~\ref{sec:base_scuc}.
From \cref{eq:generator_unavailability}, it follows that (i)  reliability of generator $g$ has no impact on the probability, if $g$ is not committed, i.e., $u_{g,t}^*=0\ \Rightarrow\ U_{g,t} = 0$, and (ii) if the generator is not starting up at time period $t$, the probability that it is unavailable is equal to its $\orr$, i.e., $v_{g,t}^*=0\ \Rightarrow\ U_{g,t} = u_{g,t}^*\orr_g$.

\subsection{Transmission Line Reliability}

The reliability of transmission lines and other interconnecting equipment, e.g., transformers, can be similarly captured via their FOR $q_{ij}$ and we define 
\begin{align}
    \orr_{ij} = 1-e^{-q_{ij}}.
\end{align}
Assuming that the topology of the network is fixed throughout the planning horizon of the SCUC model, transmission line reliability is independent of $t$ such that:
\begin{align}
    U_{ij,t} = U_{ij} = \orr_{ij},\quad \forall t\in\set{T}.
\end{align}

\subsection{Adverse Conditions}
\label{ssec:adverse_conditions}

Adverse conditions, e.g., weather effects such as extreme cold or heat, can negatively impact the reliability of generators and transmission lines \cite{ortega2016assessment}.
Equipment FORs, and subsequently ORRs, can be modified to capture increased outage rates during adverse weather conditions using the following statistical analysis. 
Let $F_g^A$ denote the share of unexpected generator outages occurring during adverse conditions. 
Further, let $H_g$ denote the total number of times for which historical data of generator $g$ is available and $H_g^A$ the number of times at which conditions are considered adverse.
As in \cite{ortega2016assessment}, we can use the following two-state model to adapt the generator FOR $q_g$ to reflect FOR during normal operation conditions $q_g^N$ and adverse operation conditions $q_g^A$ as:
\begin{IEEEeqnarray}{rlrl}
    q_g^N &= q_g \frac{H_g}{H_g - H_g^A} (1-F_g^A) &
    q_g^A &= q_g \frac{H_g}{H_g^A} F_g^A \;\;\; \\    
    \orr_g^N &= 1 - e^{-q_g^N}, &
    \orr_g^A &= 1 - e^{-q_g^A}. \;\;\;  \label{eq:adverse_orr}
\end{IEEEeqnarray}

Similar computations can be made for branch contingencies as follows:
\begin{IEEEeqnarray}{rlrl}
    q_{ij}^N &= q_{ij} \frac{H_{ij}}{H_{ij} - H_{ij}^A} (1-F_{ij}^A) &
    q_{ij}^A &= q_{ij} \frac{H_{ij}}{H_{ij}^A} F_{ij}^A \;\;\; \\    
    \orr_{ij}^N &= 1 - e^{-q_{ij}^N}, &
    \orr_{ij}^A &= 1 - e^{-q_{ij}^A}. \;\;\; 
\end{IEEEeqnarray}

\subsection{Contingency Probabilities}
\label{ssec:contingency_probabilites}

Probability $\pi_c$ of each credible contingency can be calculated as:
\begin{align}
    \pi_{c,t} = \pi_{\overline{c},t} \prod_{g\in\set{G}^c}(1 + u^*_{g,t}(U_{g,t}-1)) \prod_{ij\in\set{L}^c} U_{ij},
    \label{eq:cont_prob}
\end{align}
where:
\begin{align}
    \pi_{\overline{c},t} &=  \prod_{g \in \set{G}\setminus\set{G}^c}(1 - U_{g,t}) \prod_{ij\in\set{L}\setminus\set{L}^c}(1-U_{ij}).
\end{align}
The first term in \cref{eq:cont_prob}, $\pi_{\overline{c},t}$, represents the probability of all generators and branches that are considered operational in $\set{C}$ to operate as intended. 
The second term in \cref{eq:cont_prob} captures the probability of all generators $\set{G}^c$ to experience an outage. If a generator is not committed at time period $t$, i.e., $(U_{g,t} + (1-u^*_{g,t}))=1$ as per \cref{eq:generator_unavailability}, its reliability will not contribute to the contingency probability. 
Lastly, the third term in \cref{eq:cont_prob} represents the outage probability of all branches $\set{L}^c$.

\subsection{EENS Calculation}
\label{ssec:eens_calculation}

After solving contingency analysis \cref{mod:corrective} for each contingency scenario $c\in\set{C}$, we can calculate the EENS using load shedding results $s_{i,t,c}^{\text{LS}}$ and contingency probabilities $\pi_{c,t}$ computed in Section \cref{ssec:contingency_probabilites}. This leads to:
\begin{align}
    \eens_{c}^k &= \sum_{t\in\set{T}} \pi_{c,t}^k \sum_{i\in\set{N}} s_{i,t,c}^{\text{LS}}\\
    \eens^k &= \sum_{c\in\set{C}}\sum_{t\in\set{T}} \pi_{c,t}^k \sum_{i\in\set{N}} s_{i,t,c}^{\text{LS}},
\end{align}
where $\eens_{c}$ and $\eens^k$ capture the per-contingency and total EENS during iteration $k$, respectively.

\section{Case Study}

The case study uses the Grid Modernization Lab Consortium update of the Reliability Test System (RTS-GLMC) available at \cite{rts_glmc_git}.
The RTS-GLMC system is a 73-bus network with 120 lines. Excluding hydro generation, it hosts 73 conventional generators (nuclear, coal, gas and oil), 29 utility-scale VRES plants (Wind and PV) and 31 small-scale rooftop PV units.
Transmission line capacity has been reduced to \unit[80]{\%} and available generation from wind power plants has been reduced by \unit[40]{\%}, thus leading to the effective VRES penetration of $\approx\unit[25]{\%}$. All calculations have been performed for a 24-hour planning period using the load and VRES time series for 06/20/2020.  Table~\ref{tab:case_description} summarizes the cases with different assumptions on the underlying contingencies and their treatment, which are  compared in the case study below.
For all cases we ran the iterative process until $\eens^k \leq \unit[10^{-8}]{MWh}$ with $\lambda=0$. 
All simulations have been implemented in Python v3.8 and solved using the Gurobi Solver and Gurobi-Python API \cite{gurobi}. All experiments have been performed on a standard PC workstation with an Intel i5 processor and 16 GB RAM.

\begin{table}[]
    \centering
    \caption{Description of Cases}
    \label{tab:case_description}
    \begin{tabular}{C{0.2\linewidth}|p{0.7\linewidth}}
        \toprule
        \textbf{Case} & \textbf{Description}  \\
        \midrule
         Robust & Ensure all reserve dependent post-contingency flows as in \cref{eq:robust_deltaf}. \\
         Robust-A & As `Robust', but using adverse-condition FOR $q_g^A$, $q_{ij}^A$  to calculate contingency probabilities. See Section~\ref{ssec:adverse_conditions}. \\
         Robust-VRES & As `Robust', but wind power plants have a non-zero FOR and are considered in the contingency analysis. \\
         \midrule
         RA10 & Ensure risk-adjusted flows as in \cref{eq:riskadjusted_deltaf} with $\alpha=\unit[10]{\%}$. \\
         RA10-A & As `RA10', but using adverse-condition FOR $q_g^A$, $q_{ij}^A$  to calculate contingency probabilities. \\
         RA10-VRES & As `RA10', but wind power plants have a non-zero FOR and are considered in the contingency analysis. \\
         \bottomrule
    \end{tabular}
\end{table}

The case study uses FORs $q_g$ for conventional generators as provided in the RTS-GLMC data set, while 
failure-to-synchronize rates $q_g^s$ are taken from  \cite{ortega2008optimising} and shown in Table~\ref{tab:fos_rates}. To account for the effects of adverse weather conditions (cases `Robust-A' and `RA10-A'), it is assumed that  generators record 2 weeks per year of adverse conditions. 
Relative failure rates during adverse conditions $F_g^A$ are set to \unit[40]{\%} for all gas-fired plants, \unit[20]{\%} for all nuclear power plants, and \unit[10]{\%} for all oil- and coal-fired plants similar to  \cite{ortega2016assessment}. Also, for cases `Robust-VRES' and `RA10-VRES', we model wind farm outages. To estimate per-farm contingency probabilities, it is assumed that each wind farm consists of identical turbines each with a power rating of  $\approx$\unit[5]{MW}. 
Under this assumption,  each turbine has a FOR of \unit[8]{\%} \cite{spinato2009reliability,sulaeman2016wind}, which is used to generate a capacity outage probability table (COPT) for each farm. 
The COPT is then aggregated  into four 25-percentile bins (quartiles) to calculate the expected lost capacity in each quartile  as shown in Table~\ref{tab:wind_farm_cap_outages}. As a result, each wind farm is considered as four separate N-1 outages with a capacity loss as given in Table~\ref{tab:wind_farm_cap_outages} and an ORR of \unit[25]{\%}. 
Finally, due to the smaller scale and higher spatial distribution of PV systems, their unplanned full or partial outages never caused load shedding in our experiments. 

\begin{table}[]
\setlength\tabcolsep{4pt}
    \centering
    \caption{Failure-to-synchronize rates}
    \label{tab:fos_rates}
    \begin{tabular}{c|c c c c c c c c}
    \toprule
        Unit Group & U12 & U20 & U76 & U100 & U155 & U197 & U350 & U400 \\
        \midrule
        Fuel & oil & oil & coal & oil & coal & oil & coal & nuclear \\
        $q_g^s$ [\%] & 1.48 & 2.01 & 0.83 & 3.99 & 0.42 & 2.50 & 0.41 & 0.5 \\
    \bottomrule
    \end{tabular}
\end{table}

\begin{table}[]
    \centering
    \caption{Wind Farm Capacity Outages}
    \label{tab:wind_farm_cap_outages}
    \begin{tabular}{c|cccc}
    \toprule
    Bus \#            & 309 & 317 & 303 & 122 \\
    \midrule
    Farm MW    & 148.3        & 799.1        & 847          & 713.5        \\
    Per Turbine MW & 4.94         & 4.99         & 4.98         & 5.10         \\
    \#Turbines   & 30           & 160          & 170          & 140          \\
    \midrule
    Exp. MW loss 1st Quartile  &  0.0 & 32.1 & 63.0 & 33.0         \\
    Exp. MW loss 2nd Quartile  &  4.5 & 45.3 & 56.7 & 45.8        \\
    Exp. MW loss 3rd Quartile  &  10.8 & 57.6 & 57.5 & 53.7        \\
    Exp. MW loss 4th Quartile  &  30.3 & 110.8 & 83.3 & 81.9      \\
    \bottomrule
    \end{tabular}
\end{table}

\subsection{Convergence and Performance}
\label{ssec:eens_analysis}

\begin{figure}
    \centering
    \includegraphics[width=0.98\linewidth]{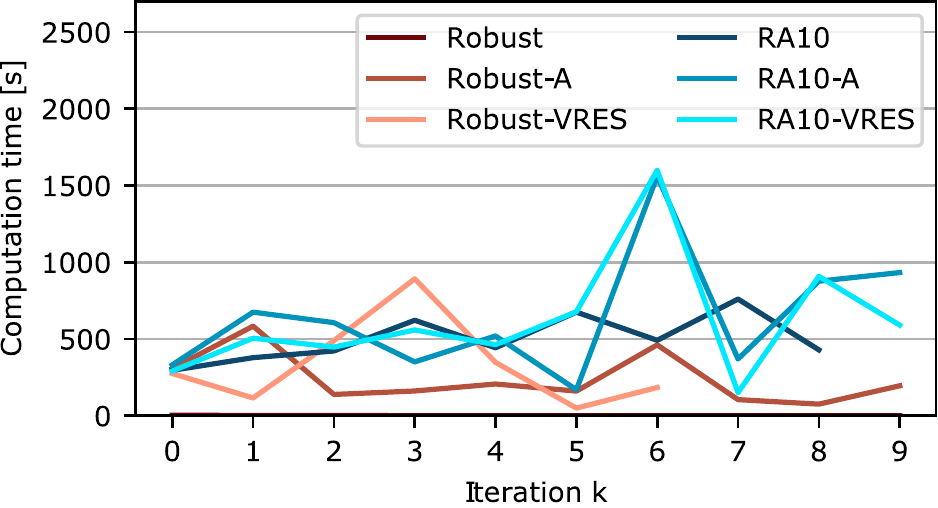}
    \caption{Computations times across iterations.}
    \label{fig:computation_times}
\end{figure}

The base SCUC was solved within \unit[386]{s}. The average solve time for the analysis of each contingency was \unit[0.14]{s}.
The solution of the ResA-SCUC required an average of {\color{black}\unit[395]{s}} with a standard deviation of {\color{black}\unit[348]{s}}.
Fig.~\ref{fig:computation_times} shows the computing  times across all iterations.
Notably, the solution times are not correlated with the number of  iterations and indicate that the proposed SCUC modifications do not systematically impact computing  times. 
We did not observe a clear correlation between the iteration number or the number of non-zero $\beta_{g,t,c}^k$ and the solve time of the ResA-SCUC.

\begin{figure}
    \centering
    \includegraphics[width=0.98\linewidth]{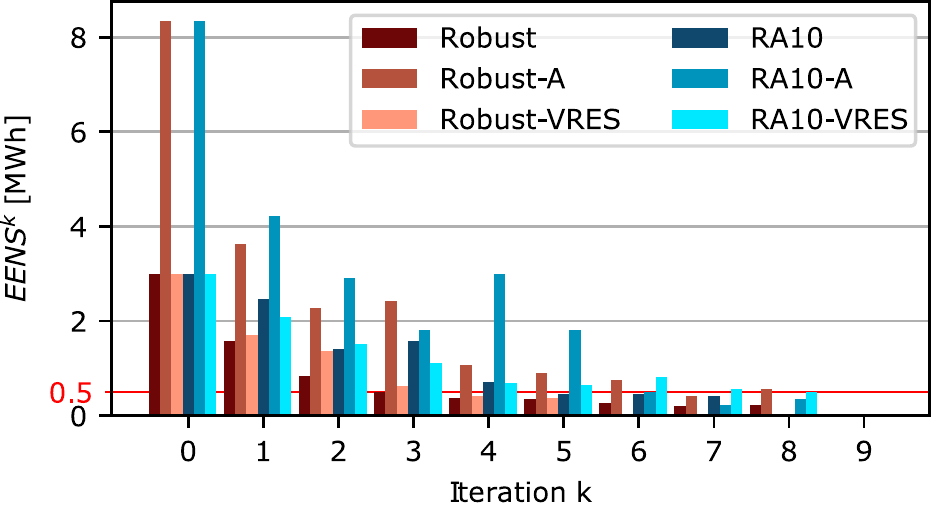}
    \caption{Comparison of $\eens^k$ at each iteration $k$.}
    \label{fig:eens_comparision}
\end{figure}

Fig.~\ref{fig:eens_comparision} shows the resulting $\eens^k$ after each iteration $k$ for all six cases in Table~\ref{tab:case_description} . 
After iteration $k=0$,  $\eens^k$ is equal for all cases but `Robust-A' and `RA10-A' because the initial SCUC calculation is the same across these  cases, which leads to the identical contingency analysis results.  Notably, the presence of VRES outages (with suffix `-VRES') does not change the  $\eens^k$, that is the considered wind farm outages do not cause load shedding. In cases  `Robust-A' and `RA10-A',    $\eens^k$ is greater due to using $q^A_g$ instead of $q^g$ for calculating $\pi_c^k$. 

With each iteration, the  $\eens^k$  reduces until it converges to zero; however, this trend is not strictly monotonic (e.g., see {\color{black}$k=3$ to $k=4$} in `RA10-A') because of the nonconvexity in \cref{mod:scuc} (i.e., binary variables). 
In all cases considered in  Fig.~\ref{fig:eens_comparision},  at most nine iterations are required. Case `Robust-VRES' terminates after  {\color{black}$k=6$} iterations, case 'RA10' -- {\color{black}$k=7$}, and cases `Robust', `Robust-A', `RA10-VRES' and `RA10-A' -- $k=9$. 
On average, the `RA10' cases result in greater $\eens^k$ values than their counterpart `Robust' cases, which is expected because the risk-adjusted flow correction in \cref{eq:riskadjusted_deltaf} ignores some less risky contingencies.

In conclusion, we observe that the proposed approach does not systematically impact computational performance. At the same time, EENS is reduced systematically and cost-effectively by determining suitable generation commitments, dispatch and reserve allocations. 

\subsection{Cost, Commitment and Reserve Analysis}
\begin{figure}[b]
    \centering
    \includegraphics[width=0.98\linewidth]{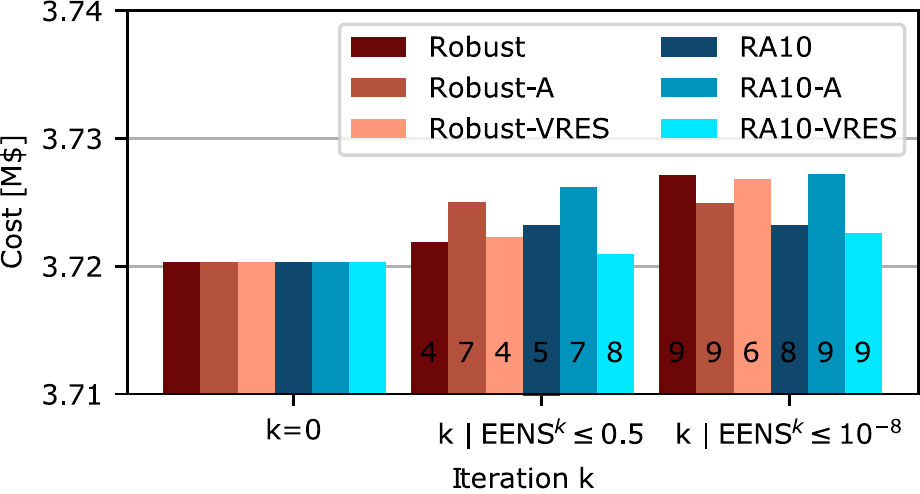}
    \caption{Total system generation cost for three iterations. The left group shows iterations $k=0$. The middle group shows the cost of the first iteration after which $\eens^k$ is below $\unit[0.5]{MWh}$ (see red line in Fig.~\ref{fig:eens_comparision}). The right group shows the cost of the first iteration after which $\eens^k$ is below $\unit[10^{-8}]{MWh}$. The iterations shown in the middle and right groups are printed in the respective bars.}
    \label{fig:cost_comparision}
\end{figure}
\begin{figure}[t]
    \centering
    \subfloat[Robust]{
        \includegraphics[width=0.95\linewidth]{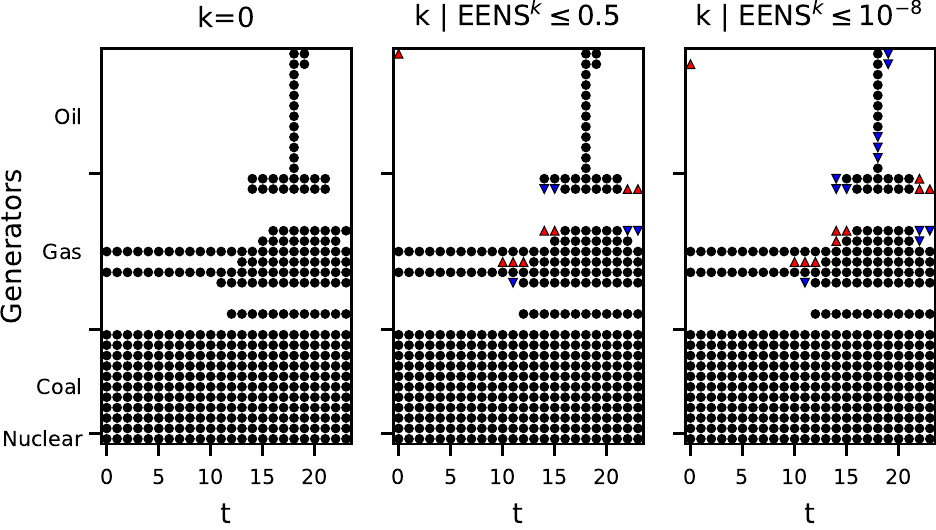}
    }\\
    \subfloat[RA10]{
        \includegraphics[width=0.95\linewidth]{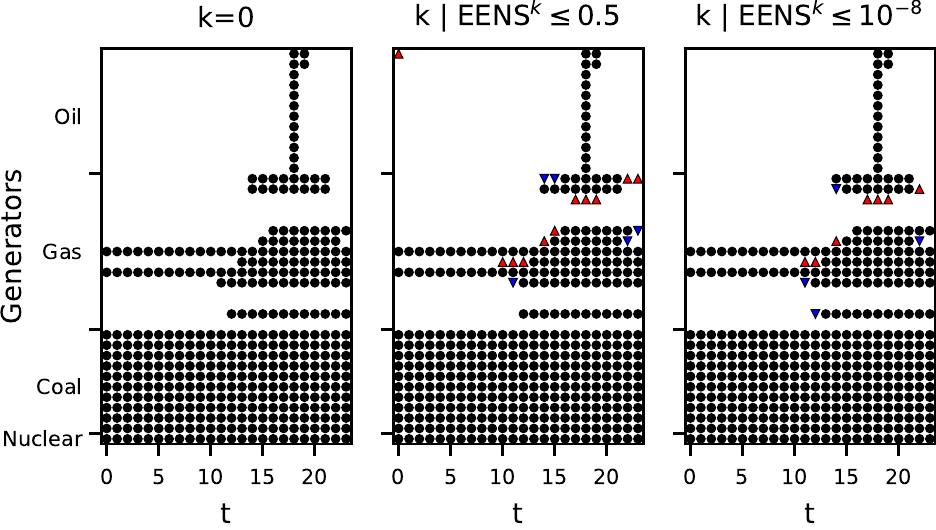}
    }
    \caption{Commitment overview for case `Robust' (a) and `RA10' (b).  Left, middle and right commitment plots correspond to iterations $0,4,8$ for `Robust' and $0,5,7$ for `RA10', i.e., the base iteration, the first iteration where $\eens^k\leq0.5$ and the first iteration where $\eens^k\leq10^{-8}$. Generators are sorted by average fuel cost from lowest (nuclear) to highest (oil). Black dots ($\bullet$) indicate that the generator is committed at time $t$. Blue downwards triangles ({\color{black}$\blacktriangledown$}) indicate that a generator that was committed in $k=0$ is not committed anymore. Red upward triangles ({\color{red}$\blacktriangle$}) indicate that a generator has been committed that was not committed in iteration $k=0$.}
    \label{fig:committments}
\end{figure}

Fig.~\ref{fig:cost_comparision}
shows the total generation cost for all cases. For  $k=0$, as discussed in Section~\ref{ssec:eens_analysis}, the initial SCUC computation is identical for all cases leading to identical cost of {\color{black}\unit[3.72]{M\$}}. As the value of  $\eens_k$  decreases to \unit[0.5]{MWh} and \unit[0]{MWh}, the operating cost increases for all cases.  
And Fig.~\ref{fig:committments} shows the resulting generator commitments for the `Robust' case in Fig.~\ref{fig:committments}(a) and for the `RA10' case in Fig.~\ref{fig:committments}(b). 
The explicit depiction of the commitment schedules for the other cases is omitted for brevity and the insights obtained from the two cases that are shown can be transferred. 

As the result of the iterative process, the total operating cost increases slightly {\color{black}($\approx 0.01 - 0.03\%$)} as the SCUC solution  deviates  from the $k=0$  commitment and  dispatch to  ensure reserve deliverability. 
Similarly to Section~\ref{ssec:eens_analysis}, the non-monotonic behavior of the total cost in some cases of  Fig.~\ref{fig:cost_comparision}, can be explained by the nonconvexity of  the SCUC optimization. Furthermore, deviations from the $k=0$ SCUC solution also affect the resulting commitment decisions. Thus, for example,  Fig.~\ref{fig:committments} shows that both the `Robust' and the `RA10' cases 
{\color{black} opt to commit additonal expensive generators for higher security levels.
This indicates that, given the generator cost curves, the SCUC would have preferred to not commit the generators indicated by the red upward triangles ({\color{red}$\blacktriangle$}), but now requires these generators to better distribute power production and reserves.
While, on the other hand, the ResA-SCUC also chooses to not commit some generators, the total number of committed generators increases for higher iteration numbers.
}

\begin{figure}[!t]
    \centering
    \includegraphics[width=0.95\linewidth]{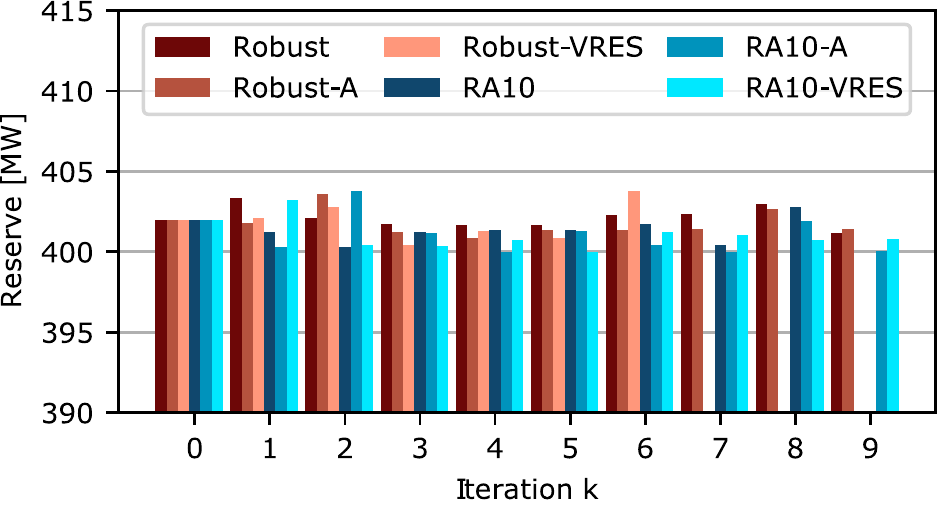}
    \caption{Total average hourly reserve for all cases and all iterations.}
    \label{fig:total_prs}
\end{figure}

\begin{figure}[!t]
    \centering
    \subfloat[Robust]{\fbox{\includegraphics[width=0.9\linewidth]{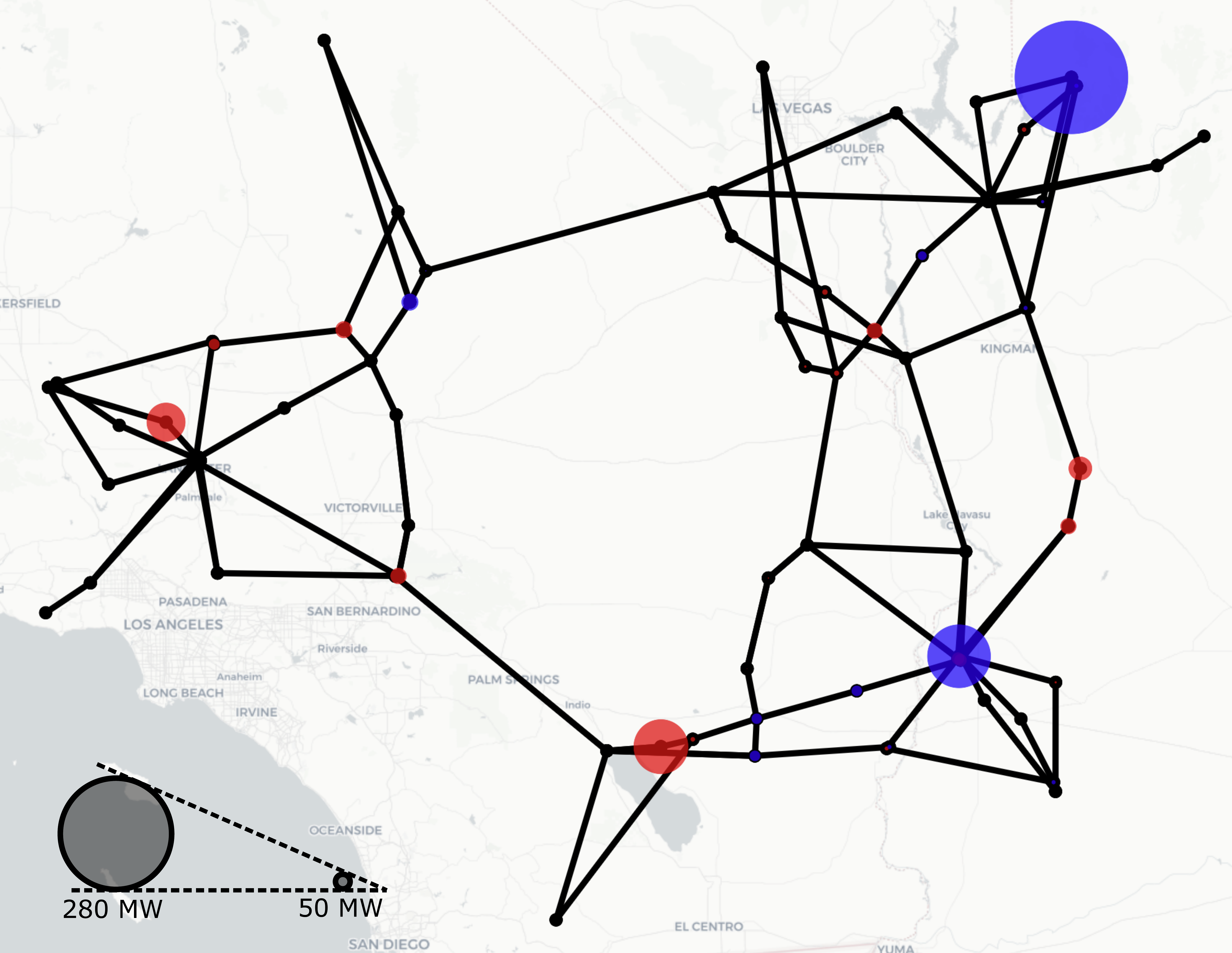}}} \\
    \subfloat[RA10]{\fbox{\includegraphics[width=0.9\linewidth]{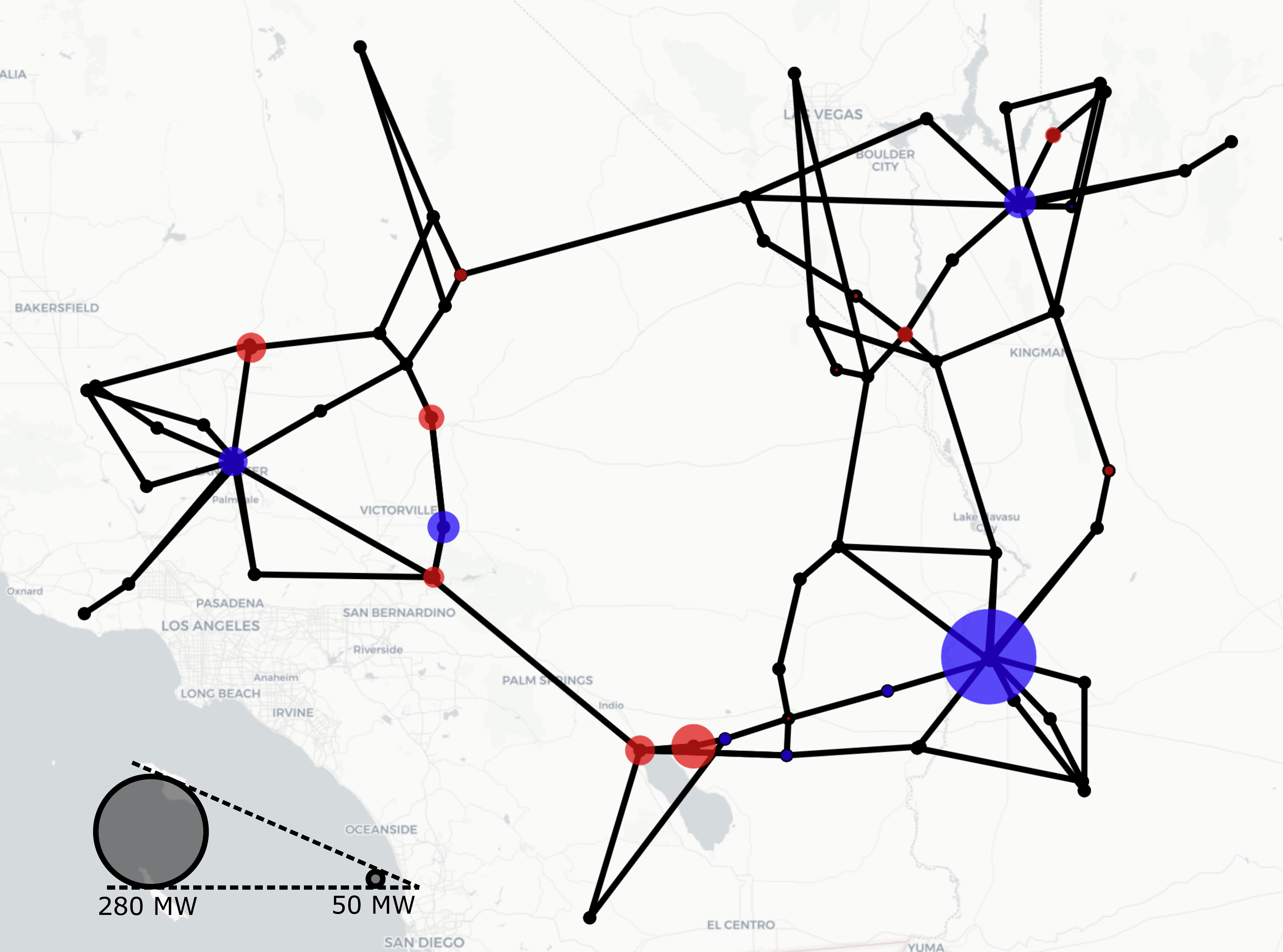}}}
    \caption{Change of allocation of total average hourly reserve between iteration $k=0$ and final iteration $k|\eens^k\leq10^{-8}MWh$ for cases `Robust' (a) and `RA10' (b).}
    \label{fig:reserve_allocaton}
\end{figure}

Fig.~\ref{fig:total_prs} itemizes the total average amount of reserve that is procured in the system for all cases and all iterations. The observed deviations from the $k=0$   SCUC solution is very small (between {\color{black}\unit[-0.6]{\%} and \unit[+0.5]{\%}}) and shows no strict correlation to the robustness of the solution, i.e., the the resulting $\eens^k$, which  indicates that reserve deliverability depends less on the total amount of procured reserves and more on the allocation of reserves in the system.
For example, Fig.~\ref{fig:reserve_allocaton} shows how the allocation of total average hourly reserve changes from iteration $k=0$ and final iteration $k|\eens^k\leq10^{-8}MWh$ for cases `Robust' and `RA10'. 
{\color{black}
In both cases, reserves are redistributed from buses with large reserve contributions to a more decentralized reserve allocation pattern.
}

\subsection{Reserve Activation Factors}
\label{ssec:reserve_activation_factors_case_study}

\begin{figure}
    \centering
    \includegraphics[width=0.98\linewidth]{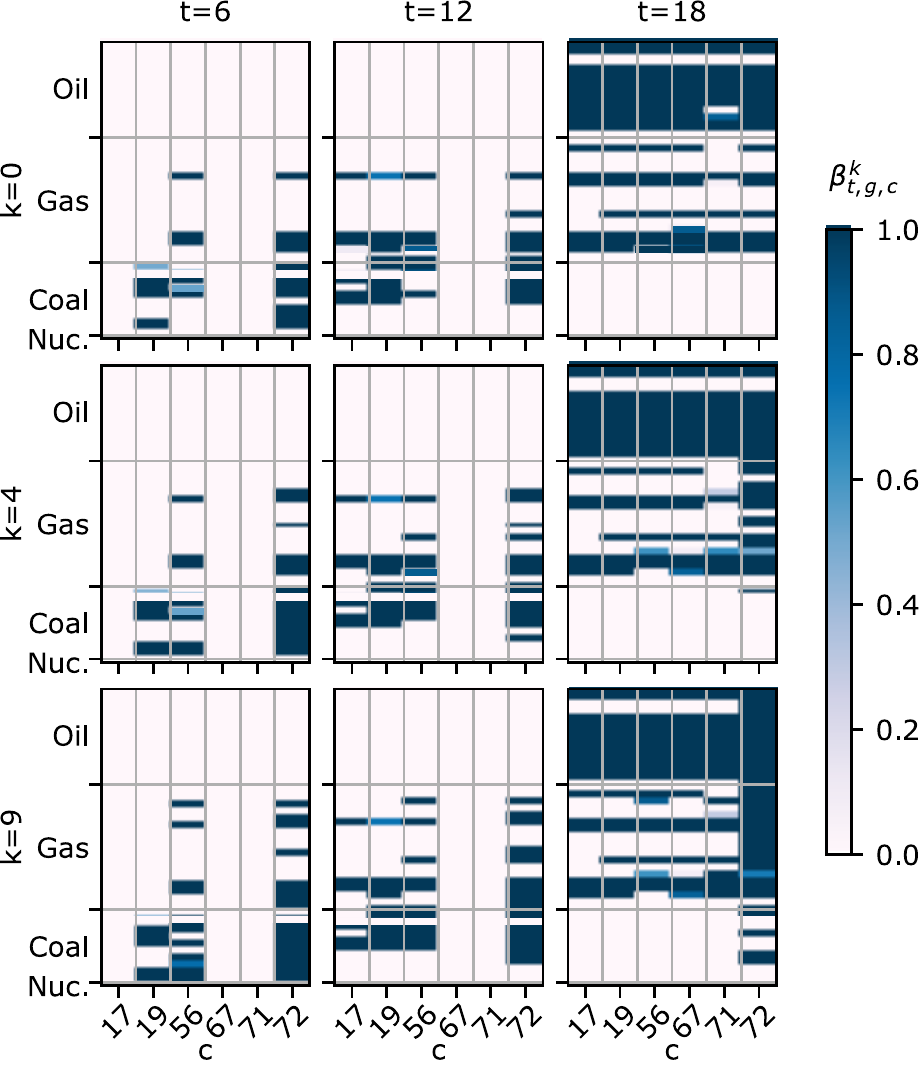}
    \caption{Values of $\beta_{g,t,c}^k$ for 'Robust' case for time steps $t\in\{6,12,18\}$ and iterations $k\in\{0,4,9\}$ for all $g$ where $\beta_{g,t,c}^k$ is non-zero at least once for any $t,c,k$. The value of $\beta_{g,t,c}^k\in[0,1]$ is indicated by the color, where a darker color indicates a higher value. Contingencies $c\in\{17, 19, 56, 67, 71, 72\}$ refer to the outage of the generator with index $c$.}
    \label{fig:betas_robust}
\end{figure}

Fig.~\ref{fig:betas_robust} illustrates reserve activation factors $\beta_{g,t,c}^k$ from the 'Robust' case for three iterations ($k\in\{0,4,9\}$, rows) and three times steps ($t\in\{6,12,18\}$, columns).
Here, generators where $\beta_{g,t,c}^k$ are always equal to zero for all $g$, $t$, and $c$ are omitted in the depiction to improve readability. 
As a result, we see that six contingencies $c\in\{17, 19, 56, 67, 71, 72\}$ are critical. Note that the index of the contingency corresponds to the index of the generator that is unavailable in that contingency. 
Most values $\beta_{g,t,c}^k$ are either equal to one (\unit[18.75]{\%} of all $\beta_{g,t,c}^k$) or equal to zero (\unit[80.05]{\%} of all $\beta_{g,t,c}^k$). Only \unit[1.20]{\%} of all $\beta_{g,t,c}^k$ are contained in $(0,1)$, indicating that only few reserves are called partially, while the majority of scheduled reserves are either called fully or not at all.

\section{Conclusion}

This paper proposed an iterative approach that allows system operators to internalize contingency risk into contingency reserve procurement. First, we demonstrate how reserve deliverability can be ensured in a standard SCUC formulation using a linear relationship between scheduled reserves and post-contingency power flows, which can be captured by reserve activation factors that are learned over time. 
Next, we use generator and transmission reliability models to compute contingency probabilities, which inform a risk assessment of the system given its scheduled reserves by means of a suitable risk metric, i.e., expected energy not served. The proposed approach maintains  computational tractability of the SCUC model and minimizes required modifications to the current SCUC practice, which streamlines its real-world adoption. 
Additionally, we have shown how contingency probabilities can be adapted to account for generator failure-to-synchronize and adverse weather conditions and demonstrated how the post-contingency power flows can be rendered risk-aware by selecting the set of worst-case contingencies for which reserve deliverability must be ensured. 
The usefulness of the proposed methodology has been shown by numerical experiments on a modified version of the Grid Modernization Lab Consortium update of the Reliability Test System.

\bibliographystyle{IEEEtran}
\bibliography{literature}

\end{document}